\documentclass[aps,pre,epsf,superscriptaddress,amsmath,amssymb,amsfonts,twocolumn,showpacs]{revtex4-1}

\usepackage{graphicx}
\usepackage{epsfig}
\usepackage{dcolumn}
\usepackage{bm}
\usepackage{braket}
\usepackage{amsmath}
\usepackage{mathtools}
\usepackage{color}
\usepackage{hyperref}
\newcommand{\abs}[1]{\left| #1 \right|} 

\usepackage{hyperref}
\usepackage{multirow}
\usepackage{bm}
\everymath{\displaystyle}
\begin{document}
%\pagewiselinenumbers
%%%%%%%%%%%%%%%%%%%%%%%%%%%%%%%%%%%%%%%%%%%%%%%%%%%%%%%%%%%%%%%%%%%%%%%%%%%%%%%
%%%%                     Title and authors                                 %%%%
%%%%%%%%%%%%%%%%%%%%%%%%%%%%%%%%%%%%%%%%%%%%%%%%%%%%%%%%%%%%%%%%%%%%%%%%%%%%%%%
\title{Pulse and continuously driven many-body quantum dynamics of\\ bosonic impurities in a Bose-Einstein condensate}

\author{K. Mukherjee}
\affiliation{Indian Institute of Technology Kharagpur, Kharagpur-721302, West Bengal, India}
\affiliation{Center for Optical Quantum Technologies, Department of Physics, University of Hamburg, 
Luruper Chaussee 149, 22761 Hamburg Germany}
\author{S. I. Mistakidis}
\affiliation{Center for Optical Quantum Technologies, Department of Physics, University of Hamburg, 
Luruper Chaussee 149, 22761 Hamburg Germany}
\author{S. Majumder}
\affiliation{Indian Institute of Technology Kharagpur, Kharagpur-721302, West Bengal, India}
\author{P. Schmelcher}
\affiliation{Center for Optical Quantum Technologies, Department of Physics, University of Hamburg, 
Luruper Chaussee 149, 22761 Hamburg Germany} \affiliation{The Hamburg Centre for Ultrafast Imaging,
University of Hamburg, Luruper Chaussee 149, 22761 Hamburg, Germany}

\date{\today}

%%%%%%%%%%%%%%%%%%%%%%%%%%%%%%%%%%%%%%%%%%%%%%%%%%%%%%%%%%%%%%%%%%%%%%%%%%%%%
%%%%%%%%%%%%%                  Abstract                        %%%%%%%%%%%%%%
%%%%%%%%%%%%%%%%%%%%%%%%%%%%%%%%%%%%%%%%%%%%%%%%%%%%%%%%%%%%%%%%%%%%%%%%%%%%%

\begin{abstract} 

We unravel the periodically driven dynamics of two repulsively interacting bosonic impurities within a bosonic bath upon considering 
either the impact of a finite pulse or a continuous shaking of the impurities harmonic trap. 
Following a pulse driving of initially miscible components we reveal a variety of dynamical response regimes depending on the driving frequency. 
At resonant drivings the impurities decouple from their host while if exposed to a high frequency driving they remain trapped in the bosonic gas. 
For continuous shaking we showcase that in the resonantly driven regime the impurities oscillate 
back and forth within and outside the bosonic medium. 
In all cases, the bosonic bath is perturbed performing a collective dipole motion. 
Referring to an immiscible initial state we unveil that for moderate driving frequencies the impurities feature a dispersive behavior whilst for 
a high frequency driving they oscillate around the edges of the Thomas-Fermi background. 
Energy transfer processes from the impurities to their environment are encountered, especially for large driving frequencies. 
Additionally, coherence losses develop in the course of the evolution with the impurities predominantly moving as a pair. 
\end{abstract}

%\pacs{67.85.−d, 67.40.Vs, 67.57.Fg, 67.57.De } 

\maketitle

%%%%%%%%%%%%%%%%%%%%%%%%%%%%%%%%%%%%%%%%%%%%%%%%%%%%%%%%%%%%%%%%%%%%%%%%%%%%%%%
%%%%%%%%%           Introduction                                  %%%%%%%%%%%%%
%%%%%%%%%%%%%%%%%%%%%%%%%%%%%%%%%%%%%%%%%%%%%%%%%%%%%%%%%%%%%%%%%%%%%%%%%%%%%%%

\section{Introduction}\label{Introduction}

%%%%%%              Section: Theoretical methods                  %%%%%%%%%%%%%
%%%%%%%%%%%%%%%%%%%%%%%%%%%%%%%%%%%%%%%%%%%%%%%%%%%%%%%%%%%%%%%%%%%%%%%%%%%%%%% 

Ultracold atoms constitute a unique testbed for monitoring the nonequilibrium quantum dynamics of strongly particle imbalanced multicomponent 
systems \cite{Will2011, Massignan2014, palzer2009,Modugno2002,Burchianti2018}. 
Recently, considerable attention has been devoted to the study of impurities in a many-body environment. 
The impurities are then dressed thereby forming quasiparticles \cite{Landau1933,Frohlich1954} such as polarons \cite{schmidt2018universal,Massignan2014}. 
Consequently, this dressing mechanism affects fundamental properties of the impurities e.g. their effective mass \cite{Khandekar1988}, 
mobility \cite{Feynman1962, Kadanoff1963}, induced interactions \cite{Massignan2014} and even allows them to form bound states known as 
bipolarons \cite{casteels2013bipolarons,camacho_2018,schmidt2018universal}. 
The exquisite tunability of the ultracold environment, e.g. the manipulation of the interaction between the impurities and their host using Feshbach resonances \cite{chin2010feshbach,kohler2006production}, 
enabled the experimental realization of both Bose \cite{Nils2016,Hu2016,catani2009entropy,fukuhara2013quantum} and Fermi polarons \cite{scazza2017repulsive,Koschorreck2012,kohstall2012metastability} and the 
consecutive probing of their characteristics. 
These include, for instance, the quasiparticle excitation spectrum via radiofrequency spectroscopy \cite{Koschorreck2012,kohstall2012metastability,cetina2015decoherence,cetina2016ultrafast}, 
the impurities trajectory employing {\it in-situ} imaging \cite{catani2009entropy,fukuhara2013quantum} and the crucial involvement of higher-order correlations \cite{koepsell2019imaging} for the 
adequate description of the polaronic states. 
Simultaneously, a vast amount of theoretical efforts have been mainly devoted to unravel the stationary properties of polaronic states \cite{grusdt2015new} 
which range from the the Fr\"ohlich model \cite{bruderer2007polaron,casteels2012polaronic,kain2016generalized,casteels2013bipolarons} to advanced beyond mean-field frameworks 
\cite{volosniev2017analytical,dehkharghani2018coalescence,Mistakidis2019,Nils2016,Ardila2018,ardila2015impurity,ardila2019analyzing,
Grusdt2018,Grusdt2017,grusdt2015new,Tempere2009,panochko2019two} that include interparticle correlations. 

Having established an adequate understanding of the stationary properties of polarons, a natural next step which has been very recently put forward 
is to investigate their nonequilibrium dynamics \cite{mistakidis2019effective,mistakidis2019quench} revealing peculiar correlation 
effects \cite{volosniev2015real,mistakidis2019quench,mistakidis2019correlated,mistakidis2019effective,Grusdt2018,shchadilova2016quantum,kamar2019dynamics,boyanovsky2019dynamics}. 
Indeed, the crucial involvement of interparticle correlations can lead to non-linear structure formation \cite{grusdt2017bose,mistakidis2019correlated}, alterations of the 
breathing frequency \cite{guebli2019effects}, the manifestation of orthogonality catastrophe phenomena \cite{mistakidis2019quench,knap2012time,Anderson1967}, dissipative motion of impurities 
in the many-body medium \cite{mistakidis2019dissipative} and also their relaxation dynamics \cite{boyanovsky2019dynamics,lausch2018prethermalization}. 
Other applications address impurity transport in optical lattices \cite{cai2010interaction,Johnson2011,Siegl2018,theel2019entanglement}, 
their collisional dynamics when penetrating with a finite velocity a gas of Tonks-Girardeau bosons 
\cite{burovski2014momentum,lychkovskiy2018necessary,meinert2017bloch,knap2014quantum,gamayun2018impact}, the effective control of quantum coherence \cite{li2019controlling} and 
investigations of three-body Effimov physics \cite{Yoshida2018, Blume2019}. 
However despite the above-mentioned first investigations, the impurities dynamics is still largely unexplored, especially in the case of more than a single impurity, 
while its further theoretical understanding is highly desirable and of growing interest with the aim to exploit this knowledge in the future for specific physical applications. 

A promising driving protocol to study the emergent nonequilibrium dynamics of impurities corresponds to a periodic driving \cite{Goldman2014, Goldman2015, Morsch2006} 
of their external potential. 
Here, the dependence of the impurities dynamical response on the driving frequency is of immediate interest in order to realize in which 
regimes \cite{mistakidis2015resonant,mistakidis2017mode,Goldman2014, Goldman2015} the impurities are dynamically trapped in their host or they can escape. 
Moreover the initial state of the system characterized as miscible when the impurities and the bath are spatially overlapping or immiscible in the opposite case 
is expected to crucially affect the dynamics. 
Another interesting prospect is to reveal induced impurity-impurity interactions \cite{dehkharghani2018coalescence,mistakidis2019induced} mediated by the environment despite the 
existence of direct $s$-wave impurity-impurity repulsions. 
Furthermore, dynamical phase separation \cite{mistakidis2018correlation,erdmann2019phase,Ao1998} and associated energy exchange \cite{nielsen2019critical,lampo2017bose} 
processes are worth studying. 
To track the driven nonequilibrium dynamics of the impurities capturing all relevant interparticle correlations we utilize the Multi-Layer Multi-Configuration 
Time-Dependent Hatree Method for atomic mixtures 
(ML-MCTDHX)~\cite{cao2017unified,cao2013multi,kronke2013non}. 
The latter is a non-perturbative variational approach especially designed to treat the correlated quantum dynamics of atomic mixtures exposed to 
time-dependent modulations. 
In particular, we consider two repulsively interacting bosonic impurities embedded in a Bose-Einstein condensate (BEC) and both being trapped in an external 
one-dimensional harmonic oscillator. 
To trigger the dynamics we apply a shaking of the harmonic potential of the impurities. 
This shaking is either performed for two driving periods and subsequently the system is left to evolve freely 
(pulse) or it is maintained throughout the evolution (continuous driving). 
The BEC is not impacted by the external driving. 
We focus on the case where the impurities and the bosonic gas are initially spatially overlapping (miscible components) while the case of an 
initially immiscible state is discussed briefly.  

Focusing on the case of a pulse and initially miscible components we unveil a variety of dynamical response regimes of the impurities 
depending on the driving frequency. 
At small driving frequencies the impurities closely follow the motion of their trap \cite{Goldman2014,Goldman2015} and after the pulse they remain 
trapped while oscillating inside the bosonic bath. 
Exposed to a resonant pulse \cite{Goldman2014,mistakidis2015resonant,mistakidis2017mode}, namely the frequency of the finite pulse is similar to the one of the harmonic trap, 
the impurities perform a complex motion escaping 
and re-entering into their host and finally decoupling from the latter. 
Entering the strongly driven regime, the impurities remain trapped in the bosonic gas and show a dispersive behavior for long evolution times. 
Considering an immiscible initial state we observe that for moderate driving frequencies the impurities feature a dispersive behavior within the bosonic gas and 
when subjected to a highly intense driving they oscillate around the edges of the Thomas-Fermi background and the intercomponent spatial separation remains intact. 

For a continuous shaking of the impurities trap and miscible components we showcase that an overall similar phenomenology 
to the non-continuously driven case occurs for very low and high driving frequencies. 
Interestingly, we observe that in the resonantly driven regime the impurities exhibit an irregular oscillatory motion, moving within and escaping 
from the BEC background, while featuring collisions with the latter. 
As a result of the impurities motion the bosonic gas is perturbed performing a collective dipole motion independently of the driving frequency 
and the protocol, a behavior that becomes more pronounced for high frequency drivings where the impurities predominantly reside within the bath. 

Moreover, we reveal that when the impurities reside within the bosonic gas they dissipate energy into the latter \cite{Mistakidis2019,mistakidis2019dissipative,nielsen2019critical,lampo2017bose}, 
a process which is more prominent for large driving frequencies where the degree of inter- and intraspecies correlations \cite{mistakidis2019correlated,mistakidis2019dissipative} is found to be enhanced. 
The development of coherence losses is unveiled by monitoring the time-evolution of the one-body coherence function \cite{li2019controlling}, while the impurities two-body reduced 
density matrix shows that they travel predominantly as a pair \cite{theel2019entanglement,mistakidis2019correlated,dehkharghani2018coalescence,mistakidis2019many,mistakidis2019induced}.  

This work is structured as follows. 
Section~\ref{theory} presents our setup and driving protocol as well as the many-body wavefunction ansatz and the 
observables which are utilized for the characterization of the periodically driven dynamics. 
In Section \ref{driving_non_cont} we discuss the emergent periodically driven dynamics of miscible components 
induced by a pulse acting on the harmonic oscillator potential of the impurities. 
The driven dynamics of initially immiscible components is showcased in Sec. \ref{immiscible_non_cont}. 
Section~\ref{continuous} presents the periodically driven time-evolution of two miscible components corresponding to a continuous shaking. 
We summarize our results and provide an outlook in Section \ref{conclusion}. 
In Appendix \ref{imbalanced_mixture} we show that the dynamical response of the mixture is not significantly affected when considering two heavy impurities. 
Appendix \ref{convergence} elaborates on the ingredients of the numerical simulations and delineates their convergence. 
Finally, Appendix \ref{shaking_single_component} showcases the dynamics of two bosons in a shaken harmonic trap.

\section{Theoretical Framework}\label{theory} 

\subsection{Hamiltonian and driving protocol}    

We consider a highly particle imbalanced bosonic mixture consisting of $N_I=2$ and $N_B=100$ atoms such that $N_B >> N_I$. 
Both species possess the same mass, i.e. $M_I = M_B = M$, and are confined in an one dimensional harmonic oscillator potential of 
frequency $\omega_I=\omega_B\equiv\omega=0.3$. 
Such a mass balanced mixture can be experimentally realized e.g. by a binary BEC of $^{87}$Rb atoms prepared in the hyperfine states 
$\Ket{F=1, m_F=-1}$ and $\Ket{F=2, m_F=1}$ \cite{egorov2013measurement}. 
The mixture is initialized in its ground state configuration (see also below) and in order to trigger the out-of-equilibrium dynamics 
the harmonic oscillator potential of the impurities is periodically shaken while the potential of the bosonic gas remains unperturbed. 
The corresponding many-body Hamiltonian reads 
\begin{equation}\label{Hamiltonian}
 \begin{split}
&H = \sum_{\sigma = B, I}^{}\sum_{i = 1}^{N_\sigma} -\frac{\hbar^2}{2M}\bigg(\frac{\partial}{\partial x_i^{\sigma}}\bigg)^2  
+\sum_{i = 1}^{N_B} \frac{1}{2} M \omega (x_i^{B})^2\\&+\sum_{i = 1}^{N_I} V_{sh}^{I}(x^{I}_i,t)  + g_{BB}\sum_{ i \geq j }^{} \delta(x^{B}_i - x^{B}_j)\\& 
+ g_{II}\sum_{ i \geq j }^{} \delta(x^{I}_i - x^{I}_j)   
+ g_{BI}\sum_{i=1}^{N_B} \sum_{j=1}^{N_I} \delta(x^{B}_i - x^{I}_j).
\end{split}
\end{equation}
Here, the periodically driven harmonic oscillator potential of the impurities takes the form 
\begin{equation}
V_{sh}^{I}(x^{I},t) = \frac{1}{2}M\omega^2 \big(x^{I} - \mathcal{A}\sin(\omega_D t)\big)^2,\label{driving_protocol}
\end{equation} 
with $\mathcal{A}$ and $\omega_D$ being the amplitude and the frequency 
of the driving respectively. 
Experimentally, this periodically driven scheme can be accomplished e.g. via acousto-optical modulators \cite{parker2013direct}. 
Moreover we operate in the ultracold regime and hence $s$-wave scattering constitutes the dominant interaction process. 
Consequently, both the intra- and the interspecies interactions are modeled by a contact potential with effective coupling constants 
$g_{BB}$, $g_{II}$ and $g_{BI}$. 
The effective one-dimensional coupling strength \cite{Olshanii1998} acquires the form 
${g_{\sigma \sigma'}} =\frac{{2{\hbar ^2}{a^s_{\sigma \sigma'}}}}{{\mu a_ \bot ^2}}{\left( {1 - {\left|{\zeta (1/2)} \right|{a^s_{\sigma \sigma'}}}
/{{\sqrt 2 {a_ \bot }}}} \right)^{ -1}}$, with $\sigma,\sigma'=B, I$, $\mu=\frac{M}{2}$ being the reduced mass and $\zeta$ is the Riemann 
zeta function. 
The transverse length scale is set by $a_\perp = \sqrt{\hbar/M\omega_\perp}$, where $\omega_\perp$ is the frequency of the transverse confinement. 
Additionally, ${a^s_{\sigma \sigma'}}$ is the three-dimensional $s$-wave intra ($\sigma=\sigma'$) or interspecies ($\sigma \neq \sigma'$) 
scattering length. 
As a result $g_{\sigma\sigma'}$ can be tuned experimentally via ${a^s_{\sigma \sigma'}}$ through Feshbach resonances \cite{kohler2006production,chin2010feshbach} 
or by manipulating ${{\omega _ \bot }}$ with the aid of confinement-induced resonances \cite{Olshanii1998}. 

For convenience, below, the many-body Hamiltonian of Eq. (\ref{Hamiltonian}) is casted in units of $\hbar \omega_{\perp}$. 
Consequently, the length, time and the interaction strength are rescaled in units of $\sqrt{ \hbar/M\omega_{\perp}}$, $\omega^{-1}_{\perp}$, and 
$\sqrt{\hbar^3 \omega_{\perp}/M}$, respectively. 
Also, the frequency $\omega$ of the harmonic oscillator and the driving frequency $\omega_D$ are expressed in terms of $\omega_{\perp}$. 
To restrict the spatial extent of the system we employ hard-wall boundary conditions at $x_\pm=\pm50$ which do not affect the dynamics 
since there is not any appreciable density population beyond $x_\pm=\pm25$. 

Our system consisting of $N_I=2$ impurities and $N_B=100$ atoms in the bosonic bath is initially prepared in its many-body ground state described by the Hamiltonian of Eq. (\ref{Hamiltonian}) 
with $\omega=0.3$, $\omega_D= 0$ and $\mathcal{A}=0$. 
Throughout this work, the intraspecies interaction strengths are kept fixed to the values $g_{II} = 0.4$ and $g_{BB} = 0.5$, unless it is stated otherwise. 
Having obtained the many-body ground state of the system with the above-mentioned parameters, we induce its nonequilibrium dynamics by considering a periodic driving of the impurities harmonic 
oscillator potential [Eq. (\ref{driving_protocol})] while the bosonic bath remains undriven. 
In particular, we employ two different driving protocols. 
Namely in the first one, which we shall term below the pulse driving, the potential of the impurities is periodically driven for only two driving periods, i.e. until $t_f=4\pi/\omega_D$, 
and afterwards the system is let to evolve freely. 
However, in the second scenario the impurities are continuously driven throughout the dynamics.    

\subsection{Many-body wavefunction ansatz}\label{wavefunction_ansatz} 

To unravel the periodically driven dynamics of the binary bosonic mixture we employ the variational ML-MCTDHX \cite{cao2017unified,cao2013multi,kronke2013non} method.  
It is based on expanding the total many-body wavefunction of the system with respect to a time-dependent and variationally optimized basis set.  
This allows us to take into account both the intra- and the interspecies correlations of the binary system using a numerically feasible size of the basis set. 
The total many-body wavefunction can be expressed in the form of a truncated Schmidt decomposition~\cite{Horodecki2009} of rank $D$ as follows 
\begin{equation} \label{wfn_ansatz}
\Psi_{MB}(\vec{x}^B, \vec{x}^I;t) = \sum_{k = 1}^{D} \sqrt{\lambda_k(t)}\Psi^B_k(\vec{x}^B;t)\Psi^I_k(\vec{x}^I;t). 
\end{equation}
The time-dependent basis states $\Psi^{\sigma}_k(\vec{x}^{\sigma};t)$ form an orthonormal $N_{\sigma}$-body wavefunction set in a subspace of the 
$\sigma$-species Hilbert space $\mathcal{H}^{\sigma}$ and are known as the species functions of the $\sigma$-species. 
Moreover, the Schmidt coefficients $\lambda_k(t)$ in decreasing order are referred to as the natural species populations of the $k$-th species function. 
These coefficients signify the presence of entanglement of the system. 
In particular, if there is only one non-vanishing Schmidt coefficient, then the total many-body state of Eq. (\ref{wfn_ansatz}) is a direct product of the two species states and the system is 
non-entangled. 
In contrast, when at least two $\lambda_k(t)$ possess a non-zero value, the system is termed entangled or interspecies correlated~\cite{Roncaglia2014}. 

Next in order to incorporate intraspecies correlations into our many-body ansatz we expand each of 
the species functions $\Psi^{\sigma}_k (\vec x^{\sigma};t)$ with respect to permanents of $d_{\sigma}$ distinct 
time-dependent single-particle functions (SPFs) $\varphi_1^{\sigma},\dots,\varphi_{d_{\sigma}}^{\sigma}$.  
Then, $\Psi^{\sigma}_k (\vec x^{\sigma};t)$ reads 
\begin{equation}
\begin{split}
&\Psi_k^{\sigma}(\vec x^{\sigma};t) = \sum_{\substack{l_1,\dots,l_{d_{\sigma}} \\ \sum l_i=N}} C_{k,(l_1,
\dots,l_{d_{\sigma}})}(t) \\& \times \sum_{i=1}^{N_{\sigma}!} \mathcal{P}_i
\left[ \prod_{j=1}^{l_1} \varphi_1^{\sigma}(x_j;t) \cdots \prod_{j=1}^{l_{d_{\sigma}}} \varphi_{d_{\sigma}}^{\sigma}(x_{K(d_{\sigma})+j};t) \right].  
\label{SPF}
\end{split}
\end{equation}  
Here, $\mathcal{P}$ is the permutation operator which exchanges the particle positions $x_{\nu}^{\sigma}$, $\nu=1,\dots,N_{\sigma}$ within the SPFs. 
Also, $K(r)\equiv \sum_{\nu=1}^{r-1}l_{\nu}$, with $l_{\nu}$ being the occupation of the $\nu$th SPF and $r\in\{1,2,\dots,d_{\sigma}\}$ and 
$C_{k,(l_1,\dots,l_{d_{\sigma}})}(t)$ are the time-dependent expansion coefficients. 
The eigenfunctions of the $\sigma$-species one-body reduced density matrix 
$\rho_\sigma^{(1)}(x,x^\prime;t)=\langle\Psi_{MB}(t)|\hat{\Psi}^{\sigma \dagger}(x)\hat{\Psi}^\sigma(x^\prime)|\Psi_{MB}(t)\rangle$ 
are termed natural orbitals $\phi^{\sigma}_i(x;t)$.   
Note that $\hat{\Psi}^{\sigma}(x)$ is the $\sigma$-species bosonic field operator. 
The natural orbitals are related with the SPFs by employing a unitary transformation that diagonalizes 
$\rho_\sigma^{(1)}(x,x^\prime;t)$ when it is expressed in the basis of SPFs, see also \cite{cao2017unified,cao2013multi,kronke2013non} for details.  
The eigenvalues of $\rho_\sigma^{(1)}(x,x^\prime;t)$ are the so-called natural populations $n^{\sigma}_i(t)$ and provide a measure 
for the occurrence of the $\sigma$-species intraspecies correlations. 
Indeed, the $\sigma$-species subsystem is intraspecies correlated if more than one eigenvalue is macroscopically occupied, otherwise 
it is termed fully coherent. 

Having specified the many-body wavefunction ansatz introduced in Eqs. (\ref{wfn_ansatz}) and (\ref{SPF}) one determines the corresponding 
ML-MCTDHX equations of motion \cite{cao2017unified,kohler2019dynamical} 
of the constituents $\lambda_k(t)$, $C_{k,(l_1,\dots,l_{d_{\sigma}})}(t)$ and $\phi^{\sigma}_j(x^{\sigma};t)$. 
Indeed, by utilizing e.g. the Dirac-Frenkel variational principle~\cite{Frenkel1934, Dirac1930} one arrives at a set of $D^2$ linear differential equations for the coefficients 
$\lambda_k(t)$, which are coupled to $D\big[(N_B + d_B -1)!/N_B!(d_B-1)! + (N_I + d_I -1)!/N_I!(d_I-1)!\big]$ non-linear integrodifferential equations for $C_{k,(l_1,\dots,l_{d_{\sigma}})}(t)$ 
and $d_I + d_B$ nonlinear integrodifferential equations for the SPFs. 
Note that for all many-body simulations, to be presented below, we use $D=10$ species functions and $d_B=3$, $d_I=6$ single-particle functions. 
In this way, numerical convergence is achieved, see Appendix \ref{convergence} for a more elaborated discussion. 
For explicit derivations and further details we refer the reader to Refs. \cite{cao2017unified,cao2013multi,kronke2013non}.  

\subsection{Observables of interest}\label{observables} 

We next introduce the main observables which will be employed for the interpretation of the periodically driven dynamics of the bosonic mixture. 
To estimate the degree of spatial one-body coherence in the course of the evolution, we invoke the normalized spatial first order 
correlation function \cite{mistakidis2018correlation, Naraschewski1999,Sakmann2008} 
\begin{align}
g^{(1)}_\sigma(x,x';t)=\frac{\rho_\sigma^{(1)}(x,x';t)}{\sqrt{\rho_\sigma^{(1)}(x;t)\rho_\sigma^{(1)}(x';t)}}.\label{one_body_coherence}
\end{align} 
Here, the $\sigma$-species one-body reduced density matrix is defined as 
$\rho_\sigma^{(1)}(x,x';t)=\langle\Psi_{MB}(t)|\hat{\Psi}^{\sigma \dagger}(x)\hat{\Psi}^\sigma(x')|\Psi_{MB}(t)\rangle$ 
whose diagonal corresponds to the one-body density $\rho_\sigma^{(1)}(x;t)\equiv\rho_\sigma^{(1)}(x,x'=x;t)$ which is accessible in cold-atom 
experiments via \textit{in-situ} imaging \cite{catani2009entropy,fukuhara2013quantum}.
Also, $\hat{\Psi}^{\sigma \dagger}(x)$ [$\hat{\Psi}^{\sigma}(x)$] is the bosonic field operator that creates [annihilates] a $\sigma$-species boson at position $x$. 
$|g^{(1)}_{\sigma}(x,x';t)|$ takes values in the interval $[0,1]$ and indicates the proximity of the many-body state to a mean-field product state for a specific set 
of spatial coordinates $x$, $x'$. 
Moreover, two distinct spatial regions $\mathcal{D}$, $\mathcal{D'}$, i.e. $\mathcal{D} \cap \mathcal{D'} = \varnothing$, are termed perfectly incoherent 
or fully coherent if $|g^{(1)}_{\sigma}(x,x';t)|= 0$ and $|g^{(1)}_{\sigma}(x,x';t)|= 1$ 
respectively with $x\in \mathcal{D}$ and $x'\in \mathcal{D'}$. 
Additionally in the case of partial incoherence, namely $0<|g^{(1)}_{\sigma}(x,x';t)|<1$, we can infer the development of one-body intraspecies correlations 
while full coherence $|g^{(1)}_{\sigma}(x,x';t)|=1$ for every $x$, $x'$ designates their absence. 

To quantify the degree of impurity-BEC interspecies correlations or entanglement during the time-evolution we exploit the so-called 
von-Neumann entropy \cite{Horodecki2009,erdmann2019phase,catani2009entropy} 
\begin{align}
S_{VN}(t)=-\sum\limits_{k=1}^D \lambda_{k}(t)\ln[\lambda_k(t)].\label{entropy} 
\end{align} 
where $\lambda_k(t)$ denote the Schmidt coefficients [Eq. (\ref{wfn_ansatz})]. 
The latter are the eigenvalues of the species reduced density matrix e.g. $\rho^{(N_{B})} (\vec{x}^{B},\vec{x}'^{B};t)=\int d x^{I} \Psi^*_{MB}(\vec{x}^{B}, 
x^{I};t) \Psi_{MB}(\vec{x}'^{B}, x^{I};t)$ where $\vec{x}^{B}=(x^{B}_1, \cdots,x^{B}_{N_{B-1}})$. 
The binary system is species entangled or interspecies correlated if more than a single eigenvalue of $\rho^{N_{B}}$ possess a non-zero value, 
otherwise it is non-entangled [see also Eq. \ref{wfn_ansatz})]. 
For instance, within the mean-field approximation $\lambda_1(t)=1$ and $\lambda_k(t)=0$, $k=2,3,\dots,D$ and therefore $S_{VN}(t)=0$, while for a 
many-body state where $\lambda_{k>1}\neq0$ it holds that $S_{VN}(t)\neq0$. 

The eigenfunctions of the $\sigma$-species one-body reduced density matrix $\rho^{(1)}_{\sigma}(x,x';t)$ are the so-called $\sigma$-species 
natural orbitals, $\phi^{\sigma}_i(x;t)$, and natural populations $\eta^{\sigma}_i(t)\in [0, 1]$ respectively. 
Each bosonic subsystem is said to be fragmented or intraspecies correlated if more than one natural population possesses a macroscopic occupation, 
otherwise the corresponding subsystem is fully coherent. 
Indeed, if the natural populations obey $\eta^{\sigma}_1(t) = 1$, $\eta^{\sigma}_{i \ne 0}(t) = 0$ [see also Eq.~(\ref{wfn_ansatz}) and Eq.~(\ref{SPF})] then 
the first natural orbital $\phi^{\sigma}_1$ becomes the Gross-Pitaevskii wavefunction $\phi^{\sigma}(x^{\sigma};t)$ \cite{kevrekidis2007emergent}. 
Accordingly, we invoke as a measure of the $\sigma$-species intraspecies correlations the deviation 
\begin{equation}
\mathcal{F}_{\sigma}(t) = 1 -\eta^{\sigma}_1(t). \label{fragmentation_measure}
\end{equation}
It provides a theoretical tool for the identification of the occupation of the $d_{\sigma}>1$ least occupied bosonic natural orbitals, 
and therefore of the deviation of the many-body from a product state when $\mathcal{F}_{\sigma}(t)>0$ \cite{Katsimiga2017,katsimiga2018many,Katsimiga_2017}. 

To monitor the position of the center-of-mass of the $\sigma$-species during the nonequilibrium dynamics we resort to 
its spatially averaged mean position 
\begin{equation}
\braket{X_{\sigma}(t)}=\braket{\Psi_{MB}(t)|\hat{x}^{\sigma}|\Psi_{MB}(t)},\label{mean_position} 
\end{equation} 
In this expression, $\hat{x}^{\sigma}=\int_{\mathcal{R}} dx x^{\sigma} \hat{\Psi}^{\sigma \dagger}(x) \hat{\Psi}^{\sigma}(x)$ represents a one-body operator and 
$\mathcal{R}$ is the spatial extension of the $\sigma$-species one-body density. 
This quantity, $\braket{X_{\sigma}(t)}$, can be assessed experimentally by relying on spin-resolved single-shot 
absorption images \cite{catani2009entropy,fukuhara2013quantum}. 
More precisely, each individual image gives an estimate of the $\sigma$-species position while $\braket{X_{\sigma}(t)}$ can be retrieved by 
averaging over several such images \cite{Katsimiga_2017,mistakidis2019dissipative,Mistakidis2018}.

\section{Driving with a two period pulse}\label{driving_non_cont} 

In this section we discuss the nonequilibrium dynamics of the bosonic mixture induced by a two period pulse driving of the harmonic 
oscillator of the impurity atoms. 
The system consisting of $N_B=100$ bosons and $N_I=2$ impurities, both being confined in the same harmonic potential of frequency $\omega = 0.3$, is 
initially prepared in its many-body ground state with $g_{BB} = 0.5$, $g_{II} = 0.4$ and $g_{BI} = 0.2$. 
Note that this choice of the interaction parameters ensures that the system initially ($t=0$) resides within the 
miscible phase since $g_{BI}<\sqrt{g_{BB}g_{II}}$~\cite{Ao1998,Timmermans1998}. 
To trigger the dynamics the harmonic oscillator potential of the impurities is periodically shaken according to Eq. (\ref{driving_protocol}) for two driving 
periods i.e. up to a finite time $t_f = 4\pi/\omega_D$ where $\omega_D$ is the driving frequency. 
Also here we consider an oscillation amplitude $\mathcal{A} = 20\gg R_{TF}$ with $R_{TF}\approx 8.3$ denoting the Thomas-Fermi radius of the bosonic gas. 
Then for $t>t_f$ the system is left to evolve freely, namely without any external driving, while keeping fixed all other parameters 
for a specific $\omega_D$. 

Moreover, we cover a wide range of driving frequencies lying in the interval $\omega_D \in [0.025, 2]$. 
Let us note in passing that the overall phenomenology, to be presented below, does not significantly depend on the value of the driving amplitude $\mathcal{A}$. 
We have verified this by inspecting the dynamics also for $\mathcal{A}=10$ and $\mathcal{A}=5$ (not shown for brevity).

\subsection{Single-particle density evolution}\label{density_non_cont}

\begin{figure*}[ht]
\centering
\includegraphics[width=\textwidth]{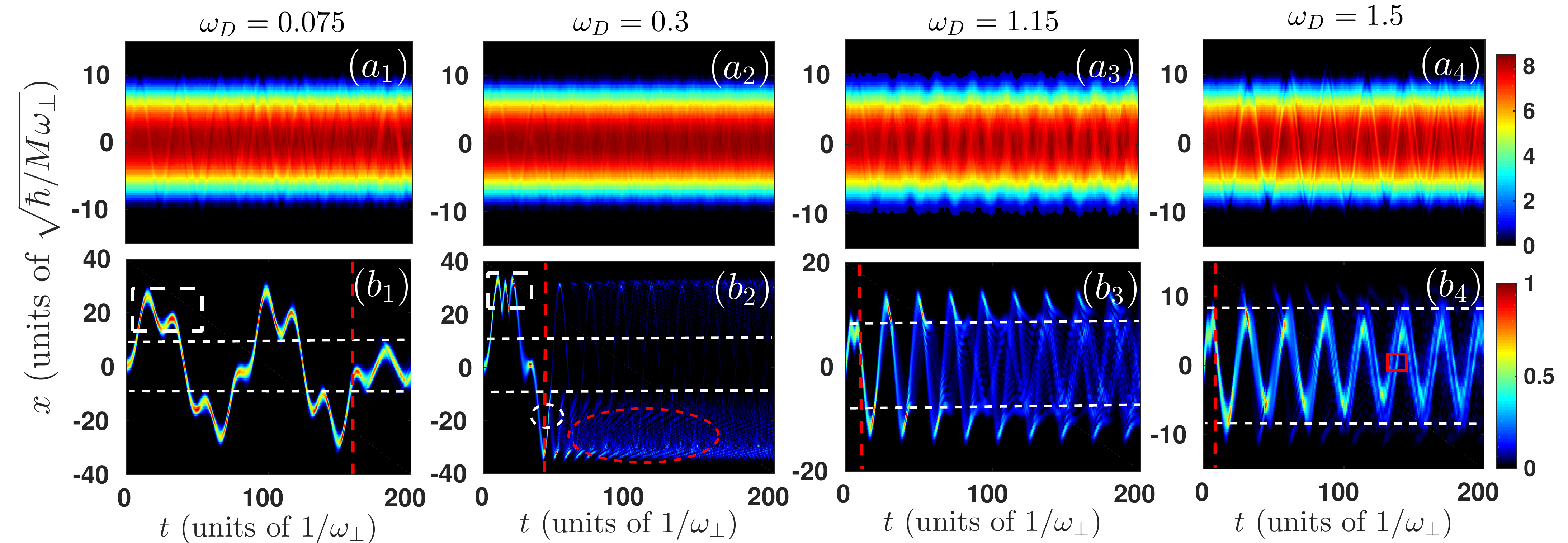}
\caption{Time-evolution of the $\sigma$-species single-particle density $\rho^{(1)}_{\sigma}(x;t)$ of ($a_1$)-($a_4$) the bosonic bath ($\sigma = B$) and ($b_1$)-($b_4$) 
the two impurities ($\sigma = I$) for specific driving frequencies $\omega_D$ (see legends). 
The shaking of the external potential of the impurities is maintained up to $t_f=4 \pi /\omega_D$ (see the dashed vertical line) and 
then the system is left to evolve. 
The bosonic mixture consists of $N_B=100$ atoms with $g_{BB}=0.5$ and $N_I=2$ interacting $g_{II}=0.4$ impurities and it is 
trapped in a harmonic oscillator of frequency $\omega = 0.3$. 
The interspecies repulsion is $g_{BI}=0.2$ and the system is prepared in its ground state. 
The dashed horizontal and vertical lines in ($b_1$)-($b_4$) indicate the location of the Thomas-Fermi radius of the bosonic gas and the time-instant ($t=t_f$) of the termination of the shaking. 
The white dashed rectangles in ($b_1$), ($b_2$) mark the back-scattering events of the impurities. 
The dashed circle and ellipse in ($b_2$) mark the splitting of $\rho^{(1)}_{I}(x;t)$ and the portion of the impurities deposited at $x<0$ respectively for $t>t_f$, while 
the rectangle in ($b_4$) indicates the density notch of $\rho^{(1)}_{I}(x;t)$. }
\label{fig:1}	 
\end{figure*}

In order to inspect the periodically driven dynamics, we first invoke the time-evolution of the $\sigma$-species single-particle density, $\rho^{(1)}_{\sigma}(x;t)$ shown 
in Fig. \ref{fig:1} for specific driving frequencies covering the low to high frequency regimes. 
For low frequencies such as $\omega_D = 0.075$, the impurities which are initially located at the trap center %[Fig. \ref{fig:2}($a_1$)] 
follow the motion of their external trapping potential moving 
towards the left edge of the bosonic gas [Fig. \ref{fig:1} ($b_1$)]. 
In this time interval $\rho^{(1)}_{I}(x;t)$ becomes more squeezed than $\rho^{(1)}_{I}(x;t=0)$ [compare the width of $\rho^{(1)}_{I}(x;t=5)$ and $\rho^{(1)}_{I}(x;t=0)$ in Fig. \ref{fig:1} ($b_1$)] 
due to the interaction of the finite velocity impurities with the BEC background. 
Then, at $t=9.2$ the impurities escape from their host reaching the position $x=27$ at $t=16$ [Fig. \ref{fig:1} ($b_1$)] which lies beyond the driving amplitude ($\mathcal{A}=20$) 
and consequently they are reflected back due to the presence of the trap. 
During this latter motion they scatter back for a short time interval when approaching the bosonic gas [see the white dashed rectangle in Fig. \ref{fig:1} ($b_1$)] 
and then reverse again their direction traveling towards the BEC which they penetrate at $t=40$. 
Note here that this weak amplitude back-scattering event is mainly caused by the driving of the harmonic trap. 
In particular, it stems from the difference between the instantaneous velocity of the impurities and the external driving at specific spatial regions.
Moreover, this event is only slightly enhanced by the presence of the repulsive $g_{BI}$, see also Appendix \ref{shaking_single_component}. 

Afterwards, the impurities dive into the bosonic bath, they reach its right edge and subsequently escape performing a similar to the above-described back and forth motion caused by the 
driven harmonic oscillator until they are again injected into the bosonic medium [Fig. \ref{fig:1} ($b_1$)].  
In this way the first oscillation of the external driving potential is completed. 
During the second oscillation period the same overall phenomenology described above occurs until $t_f=166.07$ where the driving is terminated, see the dashed vertical line in Fig. \ref{fig:1} ($b_1$). 
Subsequently, the impurities remain confined within the BEC exhibiting an oscillatory motion characterized by an amplitude of the order 
of the Thomas-Fermi radius of the bosonic gas and do not escape from the latter (see also the discussion in Sec. \ref{com_non_cont}). 
As a consequence of the impurities motion and their interaction with the BEC background $\rho^{(1)}_{B}(x;t)$ exhibits weak distortions from its original 
Thomas-Fermi profile \cite{mistakidis2019dissipative} manifested by a small amplitude collective dipole mode [Fig. \ref{fig:1} ($a_1$)] in the course of time.

Increasing the driving frequency to $\omega_D =\omega= 0.3$ (resonant driving) the impurities show a much more complex dynamical response [Fig. \ref{fig:1} ($b_2$)]. 
More precisely, at the initial stages of the dynamics they travel in the direction of the driving towards the left edge of the BEC and escape from the latter at $t=6$. 
Consequently they move until reaching $x=33.28$ where they are reflected back due to the driven harmonic trap and exhibit a back and forth motion experiencing two back-scattering 
events when approaching the boundary of the bosonic bath, see the white dashed rectangle in Fig. \ref{fig:1} ($b_2$). 
As time evolves the impurities are again injected into the bath at $t=24.8$ and interact with the latter. 

Later on, they reach the right edge of the BEC at $t=38.1$ and escape moving outwards until they arrive at the position $x=33.3$ where they feel the external oscillator 
and are reflected backwards. 
During this backward motion the external driving stops at $t=41.87$ [dashed vertical line in Fig. \ref{fig:1} ($b_2$)] and when the impurities approach the right edge of the bosonic gas 
they interact with the latter and split into two fragments [see the dashed circle in Fig. \ref{fig:1} ($b_2$)] from which one transmits through the BEC ($x>-R_{TF}$) and the 
other one is reflected back ($x<R_{TF}$). 
The reflected part [see the dashed ellipse in Fig. \ref{fig:1} ($b_2$)] possesses the major population and it remains in the right region outside of the bath throughout 
the evolution showing a dispersive behavior. 
On the other hand the transmitted portion of $\rho^{(1)}_{I}(x,t)$ performs a large amplitude oscillatory motion penetrating and escaping $\rho^{(1)}_{B}(x,t)$ during the 
time-evolution. 
Therefore after the driving the major portion of the impurities is deposited outside the right edge of the bath and as a consequence the symmetry of the population of the 
impurities with respect to the trap center is broken for $\omega_D=\omega$. 
The bosonic bath remains essentially unperturbed during the dynamics since the impurities are most of the time out of their host, see Fig. \ref{fig:1} ($a_2$). 

Next, we turn our attention to relatively fast drivings such that $\omega_D\gg \omega$ [Figs. \ref{fig:1} ($b_3$), ($b_4$)] where the time scale of the 
shaking $1/\omega_D$ is much shorter than all the other relevant time scales of the system. 
Note that in this case it is exceedingly difficult for the impurities to adjust their motion to the external periodic variation of the position of the 
trap minimum and therefore the spatial extension of $\rho^{(1)}_{I}(x,t)$ is limited to a much smaller spatial region than the driving amplitude 
$\mathcal{A}$ \cite{Goldman2014,Goldman2015,mistakidis2015resonant}. 
To support our arguments we present exemplarily the time-evolution of $\rho^{(1)}_{I}(x,t)$ corresponding to frequencies $\omega_D = 1.15$ [Fig. \ref{fig:1} ($b_3$)] 
and $\omega_D = 1.5$ [Fig. \ref{fig:1} ($b_4$)]. 
Evidently, the oscillation amplitude of $\rho^{(1)}_{I}(x;t)$ in both cases is smaller than $\mathcal{A}$. 
For instance when $\omega_D = 1.15$, $\rho^{(1)}_{I}(x;t)$ remains inside the bath within the time period of the shaking ($t<t_f=10.92$) while 
for $t>t_f$ it exhibits a periodic oscillation with an amplitude that exceeds the Thomas-Fermi radius of the BEC. 
Moreover, $\rho^{(1)}_{I}(x;t)$ has a localized shape for $t<44$ and later on it is gradually smeared out showing a relatively delocalized behavior for the time-intervals 
that the impurities lie within the bosonic gas, see Fig. \ref{fig:1} ($b_3$). 
This delocalized behavior $\rho^{(1)}_{I}(x;t)$ is caused by the interaction of the impurities with the atoms of the bath. 
However, when the impurities reside outside the edges of the BEC $\rho^{(1)}_{I}(x;t)$ forms localized spikes [Fig. \ref{fig:1} $(b_3$)]. 
The above-described dynamical response of the impurities, occurring for fast drivings, becomes more pronounced for even larger driving frequencies e.g. $\omega_D = 1.5$ 
shown in Fig. \ref{fig:1} ($b_4$). 
Indeed, $\rho^{(1)}_{I}(x;t)$ remains localized while oscillating mostly within the bosonic cloud throughout the dynamics and a decay 
of its oscillation amplitude occurs as time evolves. 
Additionally, as a consequence of the impurity-BEC interaction a density notch builds upon $\rho^{(1)}_{I}(x;t)$ for $t>105$ [see the red square box in Fig. \ref{fig:1} $(b_4)$] which 
essentially indicates the involvement of excited states in the dynamics of the impurities. 
For these fast drivings the impurities motion leave its traces on the bosonic bath manifested by the stripe patterns observed in $\rho^{(1)}_{B}(x;t)$, see Figs. \ref{fig:1} ($a_3$), ($a_4$). 
These stripe patterns reveal that the bosonic gas becomes excited due to its interaction with the impurities \cite{mistakidis2019dissipative,mistakidis2019correlated} 
and are visualized as weak distortions from its original Thomas-Fermi profile. 
Most importantly, we can deduce that the excitations of the bosonic medium are much more prominent for fast driving frequencies compared to weak ones since in the former 
case the impurities mostly reside within the majority species cloud in the course of the evolution [Figs. \ref{fig:1} ($a_3$), ($a_4$)] while in the latter case 
they escape from the BEC background [Figs. \ref{fig:1} ($a_1$), ($a_2$)]. 

A remark regarding the dynamical dressing of the impurities from the excitations of the BEC background is appropriate at this point. 
Indeed, the above-described response of the impurities to their external driving suggests that their dynamical dressing and therefore the probability to 
form a quasiparticle, here the Bose polaron, is enhanced for high driving frequencies where they mainly reside within the bosonic bath. 
However, for resonant drivings ($\omega_D=\omega$) the impurities after the termination of the driving lie outside the BEC background [Fig. \ref{fig:1} ($b_2$)] 
and as a result we can ensure that no quasiparticle is formed \cite{schmidt2018universal,Massignan2014,cetina2016ultrafast,mistakidis2019effective}.

\subsection{Time-evolution of the center-of-mass}\label{com_non_cont}

\begin{figure}[ht]
\centering
\includegraphics[width=0.45\textwidth]{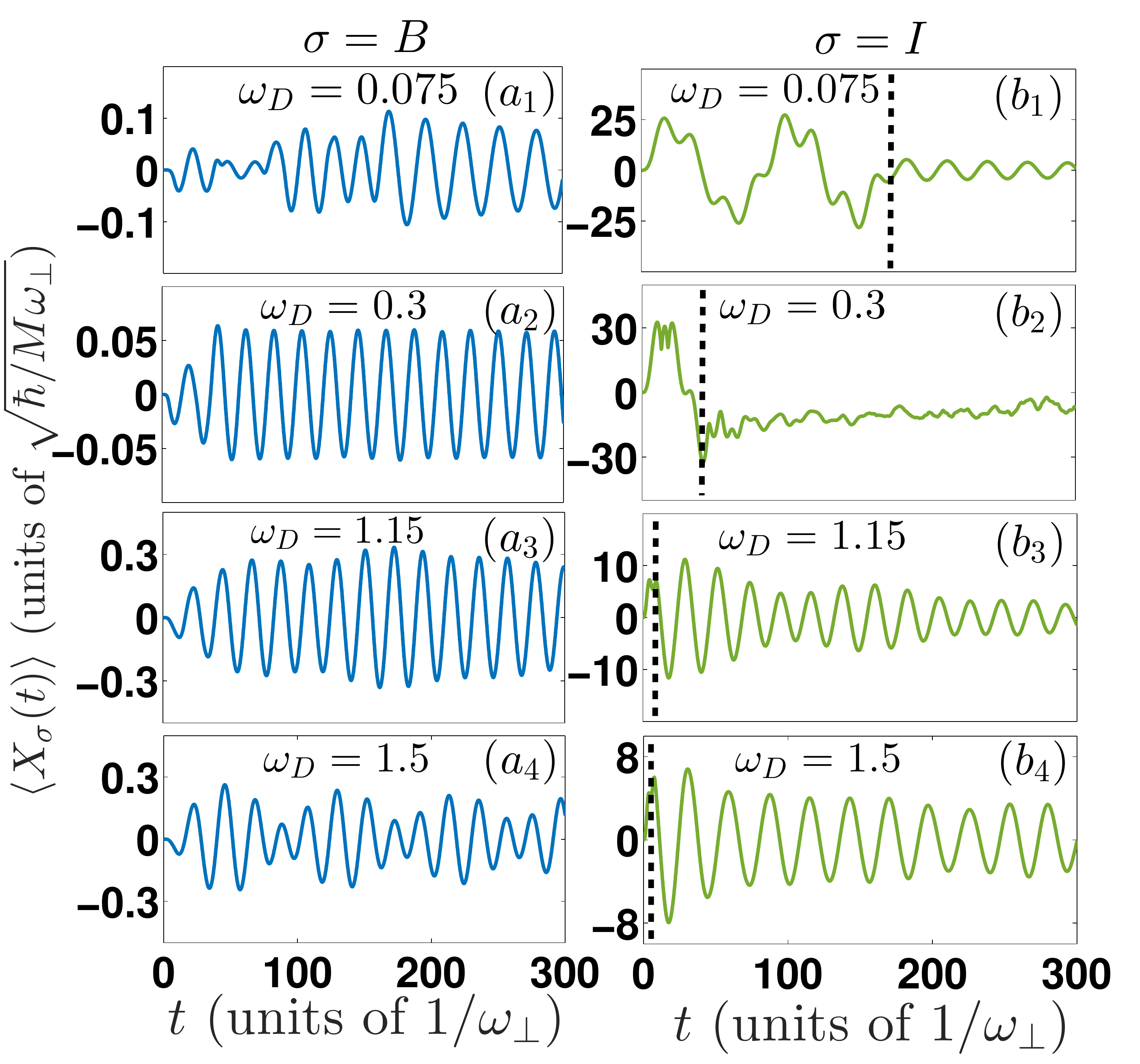}
\caption{Temporal-evolution of the center-of-mass $\langle X_{\sigma}(t) \rangle$ of the impurities ($\sigma=I$) and the bath ($\sigma = B$) 
at selective driving frequencies ($a_1$), ($b_1$) $\omega_D=0.075$, ($a_2$), ($b_2$) $\omega_D=0.3$, ($a_3$), ($b_3$) $\omega_D=1.15$ and 
($a_4$), ($b_4$) $\omega_D=1.5$. 
The dynamics is triggered by shaking the harmonic oscillator of the impurities for $t_f=4 \pi /\omega_D$ 
(see the dashed vertical lines) and then the system is left to evolve unperturbed. 
Recall that the Thomas-Fermi radius of the bosonic gas is $R_{TF}\approx 8.3$.  
All other system parameters are the same as in Fig. \ref{fig:1}.}
\label{fig:3}
\end{figure}

To gain a better understanding of the dynamical response of each species due to the periodic driving of the impurities external potential, we subsequently 
inspect the position of the $\sigma$-species center-of-mass motion captured via $\langle X_{\sigma}(t) \rangle$ [Eq. (\ref{mean_position})] for different 
$\omega_D$ [Figs. \ref{fig:3} ($a_1$)-($a_4)$ and ($b_1$)-($b_4$)]. 
Below, we first analyze $\langle X_{I}(t) \rangle$ since the impurities undergo a much more involved dynamics than the bosonic gas as argued previously. 
Recall that the impurities are directly exposed to the external driving protocol whilst the bath is only indirectly affected by their motion. 

Focusing on low driving frequencies i.e. $\omega_D = 0.075\ll\omega$ we observe that the dynamics of $\langle X_I(t) \rangle$ closely resembles 
the evolution of $\rho^{(1)}_{I}(x;t)$, compare Figs. \ref{fig:1} ($b_1$) and Fig. \ref{fig:3} ($b_1$). 
This is a consequence of the mere fact that for such low driving frequencies the impurities can adequately adapt to the externally driven trapping potential 
and therefore $\rho^{(1)}_{I}(x,t)$ remains well localized at the instantaneous trap minimum. 
In particular, for $t<t_f$, $\langle X_I(t) \rangle$ shows an ``irregular'' oscillatory behavior with an amplitude larger than the actual 
driving amplitude $\mathcal{A}=20$ [Fig. \ref{fig:3} ($b_1$)]. 
However, when the driving is terminated at $t_f=166.47$ the oscillation amplitude of $ \langle X_{I}(t) \rangle $ decreases drastically being much smaller than 
the Thomas-Fermi radius indicating that the impurities remain inside the BEC. 
Note also that in this latter time-interval the oscillation of $\langle X_I(t) \rangle$ possesses predominantly a single frequency. 
A similar ``irregular'' oscillatory pattern of $\langle X_I(t) \rangle$ takes place also for resonant drivings $\omega_D = 0.3$ when $t<t_f=41.87$ but with a comparatively 
larger oscillation amplitude than $\omega_D=0.075$, see Fig. \ref{fig:3} ($b_2$). 
Subsequently for $t>t_f$ the center-of-mass oscillation amplitude of the impurities is greatly suppressed and in particular $\langle X_I(t) \rangle$ is restricted within 
the $x<-R_{TF}\approx-8.3$ region, e.g. $\langle X_I(t=195) \rangle\approx-11$ in Fig. \ref{fig:3} ($b_2$), i.e. outside the right edge of the majority species cloud. 
Note that at $\omega_D=\omega$ and $t>t_f$ the impurities are distributed asymmetrically with respect to the bath and the major portion of their 
single-particle density is located at $x<-R_{TF}$ [Fig. \ref{fig:1} ($b_2$)]. 

In contrast to the $\omega_D=\omega$ case for fast driving frequencies e.g. $\omega_D = 1.15$ [Fig. \ref{fig:3} ($b_3$)] and $\omega_D = 1.5$ [Fig. \ref{fig:3} ($b_4$)] the 
post-shaking ($t>t_f$) time-evolution of $\langle X_I (t) \rangle$ shows a symmetric with respect to $x=0$ oscillatory pattern. 
More precisely, in both cases, the oscillation amplitude of $\langle X_I (t) \rangle$ slightly increases shortly after the termination of the shaking, e.g. $0<t<30$ in Fig. \ref{fig:3} ($b_3$), 
and then exhibits a decaying behavior in the course of time. 
Interestingly, a close comparison of $\langle X_I (t) \rangle$ between $\omega_D = 1.15$ and $\omega_D = 1.5$ reveals that the decay of its amplitude is slower in the latter case 
since for $\omega_D = 1.15$ the impurities become more delocalized within the majority species in the long time dynamics. 

Subsequently, we examine the dynamics of the center-of-mass of the BEC background, $\langle X_B(t) \rangle$, for a varying driving frequency 
illustrated in Figs. \ref{fig:3} ($a_1$)-($a_4$). 
Note that despite the fact that no external dynamical perturbation is directly applied to the majority species, the motion of the impurities is imprinted in the BEC 
as a dipole mode due to the existence of finite interspecies interactions. 
Indeed, $\langle X_B(t)\rangle$ exhibits a multifrequency oscillatory behavior independently of $\omega_D$ being characterized 
by a much smaller amplitude than $\langle X_I (t) \rangle$. 
The oscillation amplitude of $\langle X_B(t)\rangle$ acquires its smallest value for $\omega_D = \omega= 0.3$ [Fig. \ref{fig:3} ($a_2$)], 
which is attributed to the fact that for $\omega_D=0.3$ the impurities mainly reside outside the bosonic gas when the shaking is terminated, 
see also Fig. \ref{fig:1} ($b_2$).

\subsection{Interspecies energy transfer} \label{energy_non_cont}

To unveil possible energy exchange processes between the impurities and the BEC background during the periodically driven dynamics below we 
analyze the behavior of the individual intra- and interspecies energy contributions \cite{nielsen2019critical,lampo2017bose,mistakidis2019correlated}. 
The latter include the normalized or excess energy of the bath $E_B(t) = \bra{\Psi_{MB}(t) } \hat{T}_B + \hat{V}_B(x) + \hat{H}_{BB}  
\ket{ \Psi_{MB} (t)} - \bra{\Psi_{MB}(0) }  \hat{T}_B + \hat{V}_B(x) + \hat{H}_{BB}  \ket{ \Psi_{MB} (0)}$, the energy of the 
impurities $E_I(t) = \bra{\Psi_{MB}(t) }  \hat{T}_I + \hat{V}_I(x) + \hat{H}_{II}  \ket{ \Psi_{MB} (t)}$ and the interspecies 
interaction energy $E_{BI}(t) = \bra{\Psi_{MB}(t) } \hat{H}_{BI}  \ket{ \Psi_{MB} (t)}$. 
In these expressions the kinetic and potential energy operators are 
$\hat{T}_{\sigma} = -\int dx \hat{\Psi}^{\sigma \dagger}(x)\frac{\hbar^2}{2m}(\frac{d}{dx^{\sigma}})^2\hat{\Psi}^{\sigma}(x)$ and $ \hat{V}_{\sigma} 
= -\int dx \hat{\Psi}^{\sigma \dagger}(x) V^{\sigma}(x^{\sigma}, t )\hat{\Psi}^{\sigma}(x)$ respectively. 
Also, the operators of the intra- and interspecies interactions correspond to 
$\hat{H}_{\sigma \sigma} = g_ {\sigma \sigma} \int dx \hat{\Psi}^{\sigma \dagger}(x)\hat{\Psi}^{\sigma \dagger}(x)\hat{\Psi}^{\sigma}(x)\hat{\Psi}^{\sigma}(x)$ 
and $\hat{H}_{BI} = g_{BI} \int dx \hat{\Psi}^{B \dagger}(x)\hat{\Psi}^{I \dagger}(x)\hat{\Psi}^{B}(x)\hat{\Psi}^{I}(x)$.  
\begin{figure}[ht]
\centering
\includegraphics[width=0.45\textwidth]{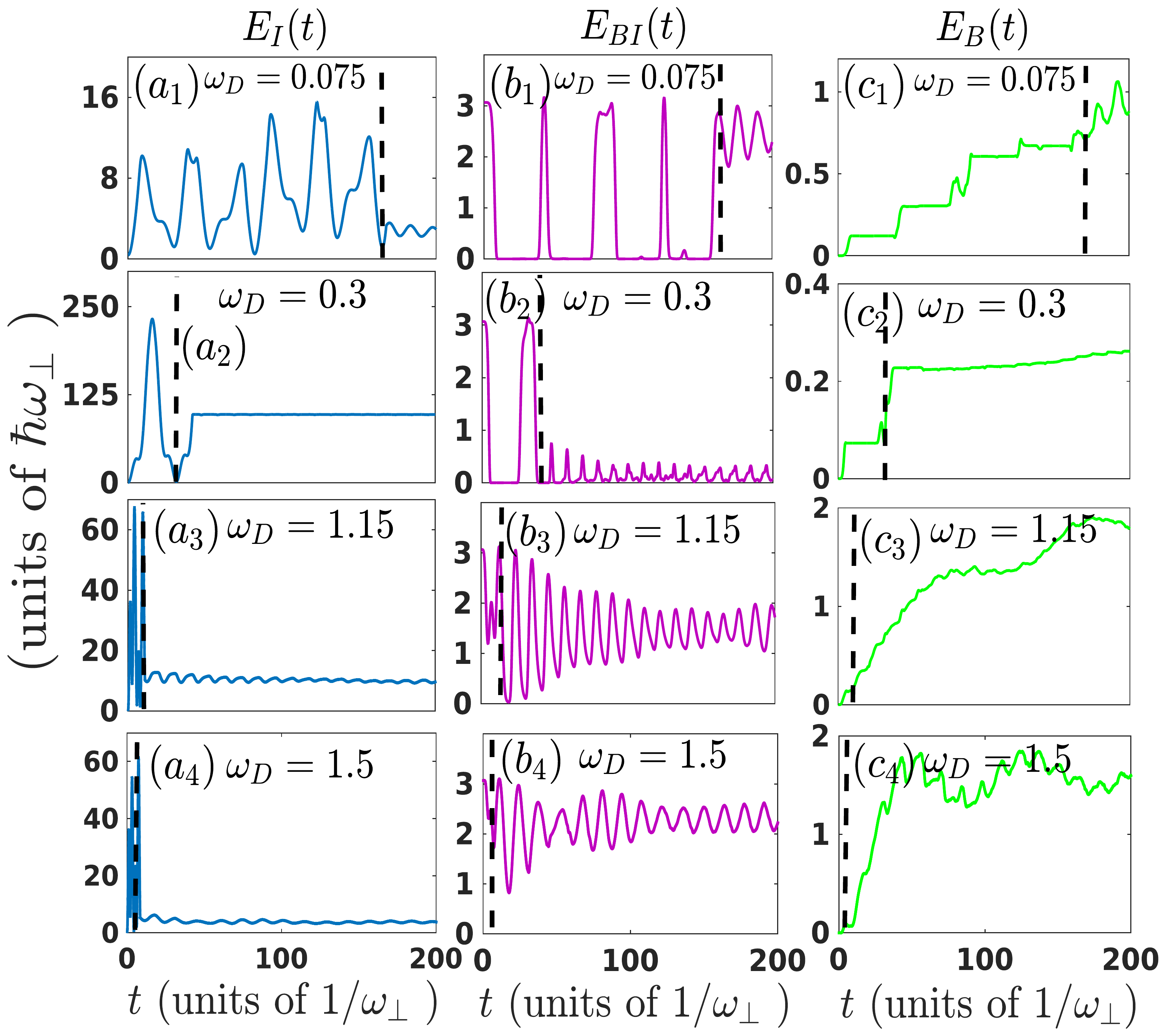}
\caption{Time-evolution of the individual energy contributions of ($a_1$)-($a_4$) the impurities $E_I(t)$, ($b_1$)-($b_4$) the interspecies 
interaction energy $E_{BI}(t)$ and ($c_1$)-($c_2$) the bosonic bath $E_{B}(t)$ at different driving frequencies $\omega_D$ (see legends). 
The dashed vertical lines mark the time-instant, $t=t_f$, of the termination of the shaking. 
The remaining system parameters are the same as Fig. \ref{fig:1}.}
\label{fig:4}
\end{figure} 

The temporal evolution of the above described energy terms is illustrated in Fig. \ref{fig:4} for distinct driving frequencies $\omega_D$. 
As it can be seen, in the course of the shaking i.e. $t<t_f$ the energy of the impurities $E_I(t)$ overall increases for every $\omega_D$ [Figs. \ref{fig:4} ($a_1$)-($a_4$)] since they 
are externally driven and therefore their kinetic energy becomes larger. 
Simultaneously the energy of the bosonic bath $E_B(t)$ exhibits an increasing tendency [Figs. \ref{fig:4} ($c_1$)-($c_4$)] while the impurity-BEC interaction 
energy $E_{BI}(t)$ decreases [Figs. \ref{fig:4} ($b_1$)-($b_4$)]. 
Notice that for low driving frequencies, namely $\omega_D=0.075$ and $\omega_D=0.3$, when $E_{BI}(t)$ tends to zero signifies that the impurities 
escape from the bosonic bath [e.g. see Fig. \ref{fig:1} ($b_1$) and Fig. \ref{fig:4} ($b_1$)] while the subsequent abrupt increase of $E_{BI}(t)$ from 
zero to a finite value is caused by the re-entering of the impurities into the bath. 
For instance at $\omega_D=0.3$ the impurities remain within the bath in the time-intervals $0\le t \le 6$ and  $ 24.8 \le t \le 38.1$ [Fig. \ref{fig:1} ($b_2$)], 
exchanging a small amount of energy with the bath [Fig. \ref{fig:4} ($b_2$)], resulting in the negligible increase of the $E_B(t)$ [Fig. \ref{fig:4}($c_2$)]. 
This behavior of the individual energy contributions suggests an energy transfer process \cite{mistakidis2019correlated,Mistakidis2019,lampo2017bose,nielsen2019critical} 
from the impurities to the bosonic gas e.g. imprinted in the density of the latter as a center-of-mass dipole mode. 
Moreover, after the application of the driving i.e. for $t>t_f$, $E_{I}$ is augmented as depicted in Figs. \ref{fig:4} ($a_1$)-($a_4$) while $E_{B}(t)$ acquires 
larger values since for $t>t_f$ the impurities reside within the BEC and thus convey energy to the latter. 
In the same time-interval the corresponding interspecies interaction energy $E_{BI}(t)$ shows an oscillatory behavior. 
In particular, when the impurities travel towards the edge of the bath [see also Fig. \ref{fig:1} ($b_2$)] they acquire more kinetic energy, 
and thus $E_{I}(t)$ increases, and transfer energy to the BEC resulting in an increase of $E_{B}(t)$ while $E_{BI}(t)$ decreases and vice versa. 
Note also that the impurities possess a maximum energy value in the case of resonant driving $\omega_D=\omega$ and as a consequence 
they are able to move far away from the Thomas-Fermi radius of the BEC [Fig. \ref{fig:1} ($b_2$)]. 

Additionally, at large driving frequencies such as $\omega_D = 1.15$ and $\omega_D = 1.5$, the energy gain of the bath is maximum since the impurities 
remain mostly within the BEC and continuously transfer energy to the latter [see Figs. \ref{fig:4} ($c_3$) and ($c_4$)]. 
For this reason also the excitations of the bath are enhanced for such large driving frequencies and its center-of-mass oscillations 
possess their largest amplitude [Figs. \ref{fig:3} ($a_3$) and ($a_4$)].

\subsection{Development of intra and interspecies correlations}\label{fragmentation_non_cont}

To estimate the degree of intra- and interspecies (entanglement) correlations in the course of the nonequilibrium dynamics of the 
bosonic mixture we resort to the deviation from unity of the first natural population $\mathcal{F}_{\sigma}(t)$ [Eq. (\ref{fragmentation_measure})] 
and the von-Neumann entropy $S_{VN}(t)$ [Eq. (\ref{entropy})] respectively. 
It is worth mentioning that $\mathcal{F}_{\sigma}(t)>0$ signifies the emergence of $\sigma$-species intraspecies correlations 
(see also Sec. \ref{observables}), while $S_{VN}(t)\neq0$ indicates the appearance of interspecies entanglement into the system \cite{mistakidis2018correlation,erdmann2019phase}. 

\begin{figure}[ht] 
	\centering
	\includegraphics[width=0.45\textwidth]{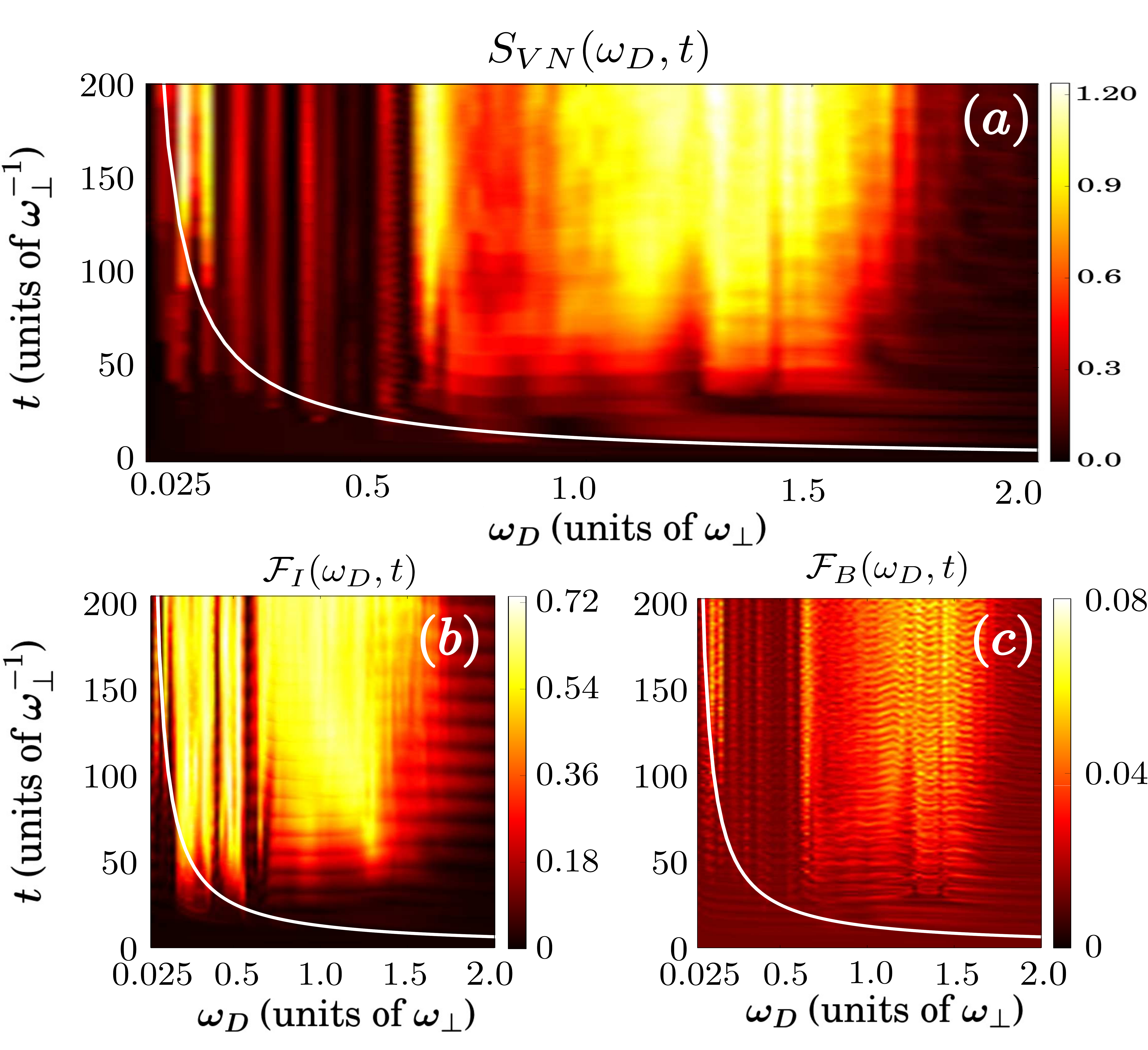}
 	\caption{Dynamics of (a) the von-Neumann entropy $S_{VN}(t)$ and the deviation from unity of the first natural population of 
 	(b) the impurities $\mathcal{F}_{I}(t)$ and (c) the bosonic gas $\mathcal{F}_{B}(t)$ for varying driving frequency $\omega_D$. 
 	The white curves indicate the duration $t=t_f=4\pi/\omega_D$ of the periodic shaking of the potential of the impurities for different 
 	driving frequencies $\omega_D$.}
 	\label{fig:5}
\end{figure}

Figure \ref{fig:5} depicts the dynamics of the von-Neumann entropy $S_{VN}(t)$ and the degree of $\sigma$-species intraspecies 
correlations $\mathcal{F}_{\sigma}(t)$ for a relevant interval of the driving frequency $\omega_D$. 
For weak and intermediate driving frequencies, i.e. $0<\omega_D<0.7$, and referring to the time-interval of the shaking 
(see the dashed white line in Fig. \ref{fig:5}) the entanglement between the species [Fig. \ref{fig:5} ($a$)] 
and the impurity-impurity correlations [Fig. \ref{fig:5} ($b$)] are small since $\max[S_{VN}(t<t_f)]<0.1$ 
and $\max[\mathcal{F}_{I}(t<t_f)]<0.12$. 
However, for later times we observe the build up of finite entanglement at particular values of $\omega_D$ and 
strong impurity-impurity correlations almost for every $0<\omega_D<0.7$. 
For instance, at $\omega_D = 0.1$, $S_{VN}(t = 10)\approx 0.03$ while $S_{VN}(t=60)\approx 0.24$ and $S_{VN}(t=180)\approx 1.15$ [Fig.\ref{fig:5} ($a$)] 
indicating the growth of entanglement via $S_{VN}(t)$ as time evolves. 
Similarly, at $\omega_D = 0.1$, $\mathcal{F}_{I}(t=10) \approx 0.01$ and $\mathcal{F}_{I}(t=60)\approx 0.07$, $\mathcal{F}_{I}(t=180)\approx 0.27$ 
[Fig.\ref{fig:5} ($b$)] designating the development of impurity-impurity correlations during the evolution. 
Recall that in the resonantly driven region i.e. $0.2 < \omega_D < 0.45$ the impurities are expelled outside the bosonic bath leading to weak interspecies 
correlations. 
Entering the high frequency driving regime, and in particular $0.75<\omega_D<1.7$, both $S_{VN}(t)$ and $\mathcal{F}_{I}(t)$ acquire larger values 
especially for $t>t_f$ indicating the existence of non-negligible interspecies entanglement and impurity intraspecies correlations. 
Note here that again in the course of the shaking, i.e. $t<t_f$, $\max[S_{VN}(t<t_f)]<0.1$ and $\max[\mathcal{F}_{I}(t<t_f)]<0.05$ implying that 
impurity-BEC and impurity-impurity correlations are mainly suppressed. 
We remark that the interspecies entanglement is maximized in the driving regime $0.75<\omega_D<1.7$ [Fig. \ref{fig:5}($a$)] since for 
such driving frequencies the impurities mostly reside within the bath, see e.g. Figs. \ref{fig:1} ($a_3$), ($b_3$), and therefore the 
impurity-BEC overlap is larger as compared to a smaller $\omega_D$. 
Turning to the driving regime $\omega_D>1.8$ we can deduce that the degree of the above-mentioned correlations almost 
vanishes since $\max[S_{VN}(t)]<0.1$ as well as $\max[\mathcal{F}_{I}(t)]<0.05$ during the entire evolution. 
This can be attributed to the fact that for such a high frequency driving the impurities cannot adapt their motion to the external 
driving thus remaining to a large extent unperturbed \cite{mistakidis2015resonant,mistakidis2017mode,Goldman2015}. 

Interestingly, by inspecting $\mathcal{F}_{B}(t)$ [Fig. \ref{fig:5} ($c$)] we can deduce that the intraspecies correlations 
of the bosonic gas are generally suppressed independently of the driving frequency. 
Indeed, only in the driving region $0.75<\omega_D<1.7$ we observe a small amount of intraspecies correlations of the bath where 
$\max[\mathcal{F}_{B}(t)]=0.07$ which occurs around $\omega_D = 1.475$. 
Otherwise it mostly holds that $\mathcal{F}_{B}(t)<0.02$ throughout the evolution. 
This behavior essentially indicates that the first natural population of the bath $\eta^{B}_1(t)$ remains close to unity 
during the time-evolution which can be partly attributed to the considered large number of $N_B=100$ particles. 

Summarizing, we deduce that for driving frequencies lying in the interval $0.75<\omega_D<1.7$ the dynamics is characterized by a significant 
amount of both intra- and interspecies correlations [Fig. \ref{fig:5}]. 
As a consequence, a many-body treatment is essential for the adequate description of the dynamics. 
However, for $0<\omega_D<0.7$ the degree of intraspecies BEC and impurity-BEC correlations is in general supressed while 
impurity-impurity correlations are finite for $t>t_f$ [Fig. \ref{fig:5}]. 
In this sense, the total many-body wavefunction and the wavefunction of the BEC can be well approximated by a mean-field product ansatz, i.e. $\lambda_1(t)=1$ in Eq. (\ref{wfn_ansatz}) and 
$n_1^B(t)=1$ in Eq. (\ref{SPF}). 
However, the wavefunction of the impurities can not be written as a product state since $n_{i>1}^I(t)>0$. 
Finally, when $\omega_D>1.8$ all correlations are mainly vanishing and therefore the driven dynamics of the system can be adequately captured 
within a corresponding mean-field treatment.

\subsection{Spatial coherence}\label{coherence_non_cont}

\begin{figure}[ht]
\centering
\includegraphics[width=0.48\textwidth]{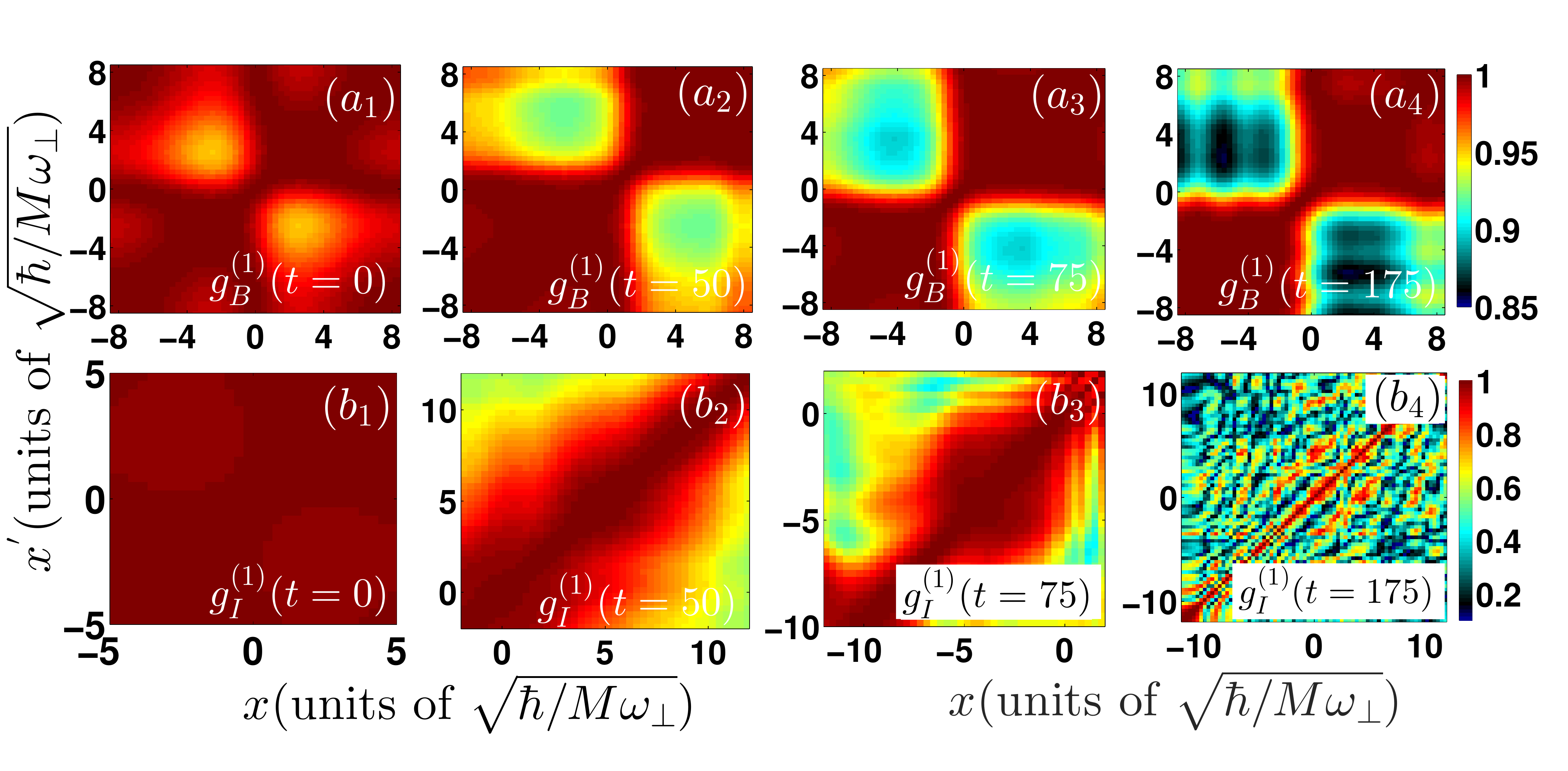}
\caption{Snapshots of the one-body coherence function $g^{(1)}_{\sigma}(x,x';t)$ of ($a_1$)-($a_4$) the bosonic bath ($\sigma=B$) and $(b_1)$-$(b_4)$ the impurities ($\sigma=I$). 
The dynamics is induced by shaking the harmonic oscillator potential of the impurities at frequency $\omega_D=1.15$ until $t=t_f=4\pi/\omega_D$ i.e. for two periods. 
All other system parameters are the same as in Fig. \ref{fig:1}.}  
\label{fig:6}
\end{figure}

To elucidate further the underlying intraspecies correlation properties of the driven bosonic mixture we investigate the $\sigma$-species one-body coherence 
function $g_{\sigma}^{(1)}(x,x';t)$ \cite{Katsimiga_2017,mistakidis2018correlation,Naraschewski1999} introduced in Eq. (\ref{one_body_coherence}). 
Recall that the situation with $g_{\sigma}^{(1)}(x,x';t)=1$ signifies that the $\sigma$-species many-body state is identical to a mean-field product ansatz. 
Therefore, if $g_{\sigma}^{(1)}(x,x';t)<1$ indicates the necessity of a beyond mean-field treatment of the dynamics. 
Figure \ref{fig:6} ($a_1$)-($a_4$) and ($b_1$)-($b_4$) present $g^{(1)}_{B}(x, x^{'}, t)$ and $g^{(1)}_{I}(x, x^{'};t)$ respectively for selected time-instants 
of the pulse driven dynamics with $\omega_D=1.15$. 
Note here that we focus on large driving frequencies since the degree of correlations is enhanced in this region as we have 
identified in the previous section \ref{fragmentation_non_cont}, see also Fig. \ref{fig:5}.

Regarding the bosonic gas, we observe that partial (i.e. very limited) losses of coherence occur between its right ($x>0$) and left ($x>0$) spatial regions since 
the off-diagonal elements of the coherence function take values smaller than unity, i.e. $g^{(1)}_B(x,x'\neq x;t)<1$ throughout the 
evolution, see Figs. \ref{fig:6} ($a_1$)-($a_4$). 
Even the initial state of the bath is not perfectly coherent see e.g. $g^{(1)}_B(x=4,x'=-4;t=0)\approx 0.96$ in Fig. \ref{fig:6} ($a_1$). 
Moreover as time evolves the aforementioned losses of coherence become more prominent, for instance $g^{(1)}_{B}(x =4.0,x'=-2.5;t = 50) \approx 0.91$ [Fig. \ref{fig:6} ($a_2$)] 
and $g^{(1)}_{B}(x=4.0,x'=-2.5;t = 175)\approx 0.88$ [Fig. \ref{fig:6} ($a_4$)] but remain very limited. 
It is also worth mentioning that the development of coherence losses in the bath during the dynamics is caused in part by the motion and interaction of the 
impurities within the BEC since the latter is not directly affected by the driving. 
Turning to the impurities we can deduce that initially small coherence losses are present between the edges of their cloud,  
see e.g.  $g^{(1)}_{I}(x=4, x'=-4;t=0)\approx0.95$ in Fig. \ref{fig:6} ($b_1$). 
In the course of the time-evolution the shaking introduces a large amount of coherence losses \cite{li2019controlling}, e.g. $g^{(1)}_{I}(x=2.8,x'=-4.5;t=50) \approx 0.80$ in Fig. \ref{fig:6} ($b_2$), 
which become substantial deeper in the evolution. 
The latter can be directly inferred from the vanishing tendency of the off-diagonal elements of the one-body coherence function. 
Indeed, $g^{(1)}_{I}(x=2.8,x'=-4.5;t=75) \approx 0.57$ [Fig. \ref{fig:6} ($b_3$)] and 
$g^{(1)}_{I}(x=2.8,x'=-4.5;t = 175) \approx 0.39$ [Fig. \ref{fig:6} ($b_4$)]. 
We remark that the spatial fluctuations of $g^{(1)}_{I}(x,x'\neq x;t)$ at long evolution times [e.g. at $t=175$ in Fig. \ref{fig:6} ($b_4$)] are pronounced 
due to the delocalized shape of the corresponding $\rho^{(1)}_{I}(x;t)$, see Fig. \ref{fig:1} ($b_3$).
Note also that the amount of coherence losses of the impurities is significantly larger than the corresponding ones occurring for the bosonic gas.

\subsection{Effect of the intraspecies interaction of the BEC on the dynamics}\label{bath_interaction}

\begin{figure}[ht]
\centering
\includegraphics[width=0.5\textwidth]{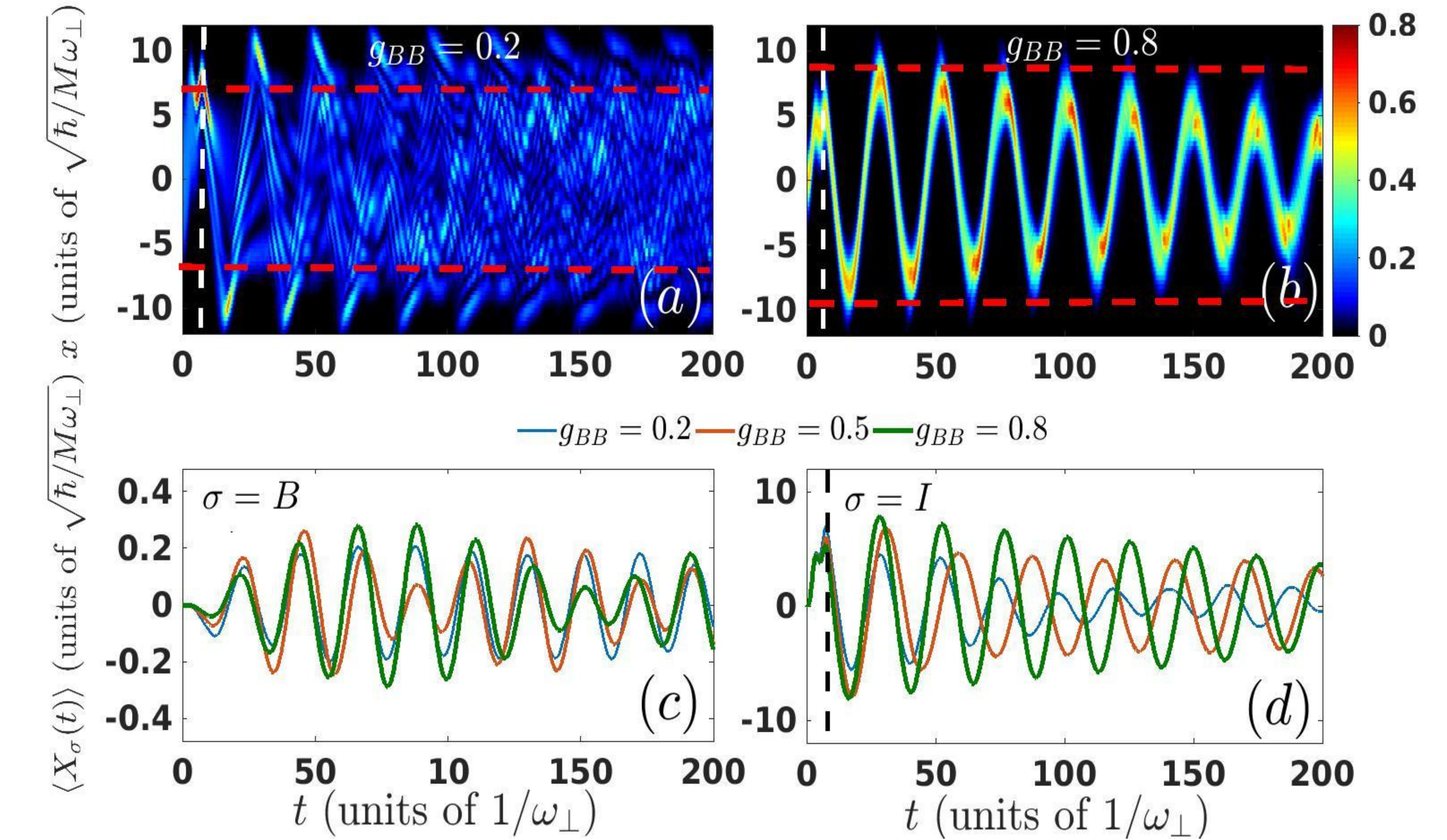}
\caption{Single-particle density evolution of the two periodically driven impurities with $\omega_D=1.5$ for ($a$) $g_{BB}=0.2$ and ($b$) $g_{BB}=0.8$. 
Dynamics of ($c$) $\langle X_{B}(t) \rangle$ and ($d$) $\langle X_{I}(t) \rangle$ 
for different intraspecies interaction strengths $g_{BB}$ of the bath (see legend) when $\omega_D=1.5$. 
The external driving of the harmonic oscillator of the impurities is performed up to $t_f=4 \pi /\omega_D$ i.e. for two periods (see the dashed vertical line). 
The mixture consists of $N_B=100$ bosons with $g_{BB}=0.5$ and $N_I=2$ interacting $g_{II}=0.4$ impurities being trapped in a harmonic 
oscillator of frequency $\omega = 0.3$. 
The interspecies repulsion is $g_{BI}=0.2$ and the system is initialized in its ground state. 
The dashed horizontal lines in ($a$)-($b$) mark the location of the Thomas-Fermi radius of the bosonic gas.}
\label{fig:7}
\end{figure}

To infer whether the intraspecies interaction of the bosonic bath can alter the above-described pulse driven dynamics we next 
investigate the dynamical response of the system for a specific $\omega_D$ and different values of $g_{BB}$. 
For simplicity, we focus on large driving frequencies e.g. $\omega_D=1.5$ where the impurity dynamics shows a more regular behavior 
[Fig. \ref{fig:1} ($b_4$)] compared to a smaller $\omega_D$, see for instance Fig. \ref{fig:1} ($b_2$), and also the degree 
of correlations is enhanced [Fig. \ref{fig:5}]. 
Figure \ref{fig:7} presents the resulting time-evolution of the impurities single-particle density $\rho^{(1)}_{I}(x;t)$ 
for $g_{BB}=0.2$ [Fig. \ref{fig:7} ($a$)] and $g_{BB}=0.8$ [Fig. \ref{fig:7} ($b$)] for $\omega_D=1.5$. 
As it can be seen, for a weakly interacting BEC background the impurities show a relatively dispersive behavior identified by 
the highly delocalized shape of $\rho^{(1)}_{I}(x;t)$ which becomes very prominent deep in the evolution, see Fig. \ref{fig:7} ($a$). 
However upon increasing the intraspecies interaction of the bath, $\rho^{(1)}_{I}(x;t)$ undergoes a decaying amplitude oscillatory motion 
within $\rho^{(1)}_{B}(x;t)$ while possessing a localized shape throughout the evolution as illustrated in Fig. \ref{fig:7} ($b$). 
Recall that this latter behavior of the impurities for $g_{BB}=0.8$ persists also when $g_{BB}=0.5$ [Fig. \ref{fig:1} ($b_4$)]. 
For weak repulsive interactions such as $g_{BB}=0.2$ and of course the same $N_B$ the bosonic bath is more dense and its Thomas-Fermi 
radius is smaller than for a stronger $g_{BB}$. 
Accordingly the interatomic distance for a weak $g_{BB}$ is smaller compared to the case of a strong $g_{BB}$ and therefore the impurities 
experience more scattering events with the atoms of the bath, resulting in the observed dispersive behavior of $\rho^{(1)}_{I}(x;t)$. 

In order to further elucidate the dynamical response of the binary system for distinct values of $g_{BB}$ we also inspect the underlying 
center-of-mass motion of both the bosonic bath $\langle X_B(t) \rangle$ [Fig. \ref{fig:7} ($c$)] and the 
impurities $\langle X_I(t) \rangle$ [Fig. \ref{fig:7} ($d$)]. 
Regarding the impurities we observe that $\langle X_I(t) \rangle$ performs a decaying amplitude single-frequency oscillatory motion independently 
of $g_{BB}$. 
Additionally, the oscillation amplitude of $\langle X_I(t) \rangle$ is smaller and its decay is more dramatic for a decreasing 
$g_{BB}$, see Fig. \ref{fig:7} ($d$). 
Indeed, the Thomas-Fermi radius of the BEC reduces for a smaller $g_{BB}$, e.g. $R_{TF}\approx 6.5$ for $g_{BB}=0.2$ and $R_{TF}\approx 9.5$ when $g_{BB}=0.8$. 
As a consequence the oscillation amplitude of $\langle X_I(t) \rangle$ is smaller for a decreasing $g_{BB}$ since for such high frequency drivings 
the impurities oscillate within the bosonic bath whose size becomes smaller. 
On the other hand, the center-of-mass of the bosonic bath as already explained in Secs. \ref{density_non_cont} and \ref{com_non_cont} undergoes an 
irregular oscillatory behavior due to the collective dipole mode caused by the motion of the impurities inside the bath. 
Here we can infer that the oscillation amplitude of $\langle X_B(t) \rangle$ is mainly increased for a stronger 
$g_{BB}$ as in the case of $\langle X_I(t) \rangle$, see Fig. \ref{fig:7} ($d$). 
However this behavior is not valid in general and notable exceptions occur during the evolution. 

\section{Pulse driven dynamics of immiscible components}\label{immiscible_non_cont}

\begin{figure}[ht]
\centering
\includegraphics[width=0.5\textwidth]{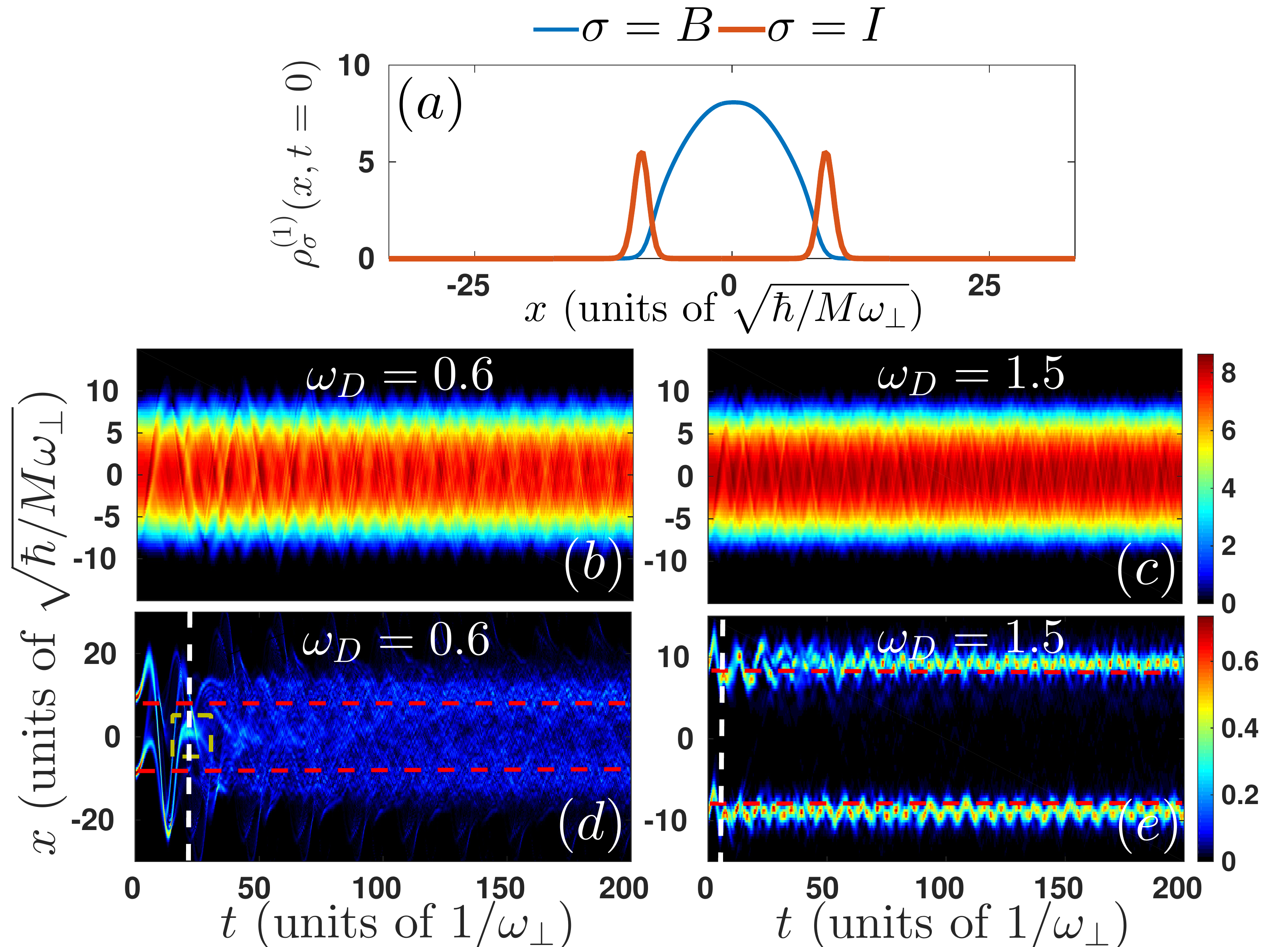}
\caption{(a) Ground state single-particle density profiles $\rho^{(1)}_{\sigma}(x)$ of the bosonic gas ($\sigma = B$) and the impurities ($\sigma = I$). 
Evolution of (b), (c) $\rho^{(1)}_{B}(x;t)$ and (d), (e) $\rho^{(1)}_{I}(x;t)$ for different driving frequencies (see legends). 
The periodic driving of the impurities harmonic oscillator is maintained up to $t_f=4 \pi /\omega_D$ (see the dashed vertical lines) and 
then the system evolves freely. 
The harmonically trapped ($\omega = 0.3$) bosonic mixture comprises of $N_B=100$ atoms and $N_I=2$ impurities with intra- and interspecies repulsions $g_{BB}=0.5$, $g_{II}=0.4$ 
and $g_{BI}=1.0$ respectively and it is prepared in its ground state. 
The dashed horizontal lines in ($d$), ($e$) indicate the location of the Thomas-Fermi radius of the bosonic gas while the grey dashed rectangle in ($d$) marks the bunching of the 
impurities around $x=0$.}  
\label{fig:14}
\end{figure}

Having analyzed in detail the driven dynamics of two miscible species whose interactions satisfy the condition 
$g^2_{BI} \le g_{II}g_{BB}$ \cite{Ao1998, Timmermans1998} we then discuss the corresponding nonequilibrium dynamics 
initializing the binary system in an immiscible state. 
As in Sec. \ref{driving_non_cont}, the highly imbalanced mixture comprises of $N_B=100$ bosons and $N_I=2$ impurities 
and both species are trapped in the same harmonic potential with $\omega = 0.3$. 
In order to realize an immiscible initial configuration we consider the same intraspecies interactions as before, namely 
$g_{BB} = 0.5$ and $g_{II} = 0.4$, but a stronger interspecies repulsion $g_{BI} = 1.0$ such that the immiscibility 
condition $g^2_{BI} \ge g_{II}g_{BB}$ is satisfied. 
The system is initially prepared into its many-body ground state where the single-particle densities of the individual species are spatially 
phase separated as shown in Fig. \ref{fig:14} (a). 
In particular, $\rho^{(1)}_{B}(x)$ resides around the trap center while $\rho^{(1)}_{I}(x)$ forms a two density hump structure with each hump 
located at an edge of the Thomas-Fermi radius, $R_{TF}\approx 8.3$, of the bosonic gas. 

The nonequilibrium dynamics is induced by considering a pulse shaking of the impurities harmonic oscillator described by 
Eq. (\ref{driving_protocol}) where the system is periodically driven for two driving periods, $t_f = 4\pi/\omega_D$, and then is 
left to evolve without any external perturbation. 
The driving amplitude is $\mathcal{A} = 20\gg R_{TF}$. 
For simplicity we study below the cases of large driving frequencies namely $\omega_D=0.6$ and $\omega_D=1.5$ since for weak $\omega_D<0.5$ 
the resulting dynamics of the impurities is found to be highly dispersive i.e. $\rho^{(1)}_{I}(x;t)$ exhibits a strongly 
delocalized behavior within and outside the BEC background (results not shown here for brevity). 
Note also that the driving frequencies $\omega_D=0.6$ and $\omega_D=1.5$ are representative of the so-called 
intermediate and high frequency driving regimes respectively. 

Figures \ref{fig:14} (b), (d) and (c), (e) depict the time-evolution of the $\sigma$-species single-particle density following a 
periodic driving with $\omega_D=0.6$ and $\omega_D=1.5$ respectively. 
For $\omega_D=0.6$ and referring to $t<t_f=20.9$ (see the dashed line in Fig. \ref{fig:14} (d)) we observe that both density humps 
of $\rho^{(1)}_{I}(x;t)$ ensue their external trap until it reaches for the first time its maximum displacement $\mathcal{A}$. 
Then as the harmonic oscillator turns towards $x=0$ the density humps collide while emitting small amplitude density fragments due to their 
interaction with the atoms of the bath and at $t=t_f$  $\rho^{(1)}_{I}(x;t)$ is predominantly concentrated at the origin $x=0$, see the grey dashed rectangle in Fig. \ref{fig:14} ($d$).  
Subsequently, for $t>t_f$, the impurities are predominantly trapped into $\rho^{(1)}_{B}(x;t)$ throughout the dynamics and therefore the species are completely mixed. 
In particular, $\rho^{(1)}_{I}(x;t)$ shortly after $t_f$ exhibits a delocalized behavior [Fig. \ref{fig:14} (d)] inside $\rho^{(1)}_{B}(x;t)$ while for $t>120$ it shows a 
tendency to segregate into two fragments symmetrically placed around $x=0$ and being located around the edges of the BEC background [Fig. \ref{fig:14} ($b$)]. 
Moreover the motion of the impurities within $\rho^{(1)}_{B}(x;t)$ perturb the bath which in turn performs a collective dipole motion [Fig. \ref{fig:14} (b)]. 
It is worth noticing here that besides the inherent tendency of the two species to remain spatially separated due to the strong $g_{BI}$, the driving enforces the 
impurities to infuse into the BEC in the course of dynamics [Fig. \ref{fig:14} ($d$)]. 

In contrast to the above-described mixing dynamics, a sufficiently high frequency driving e.g. $\omega_D = 1.5$ [see Figs. \ref{fig:14} ($c$), ($e$)] 
preserves the phase separation in the course of the evolution even after $t=t_f= 8.37$. 
Indeed, the initial density humps of $\rho^{(1)}_{I}(x;t)$ remain while oscillating at the edges of the bosonic gas throughout the dynamics. 
Interestingly, the structures building upon each density hump of $\rho^{(1)}_{I}(x;t)$ are not identical, for instance $\rho^{(1)}_{I}(x)$ is fragmented 
in the upper ($x>0$) hump within $22<t<45$ but not in the lower ($x<0$) one [Fig. \ref{fig:14} ($e$)]. 
This difference is caused by the location of each density hump at $t=t_f$. 
Indeed the upper hump at $t=t_f$ resides within the bath and therefore it interacts with the latter while the lower hump lies 
at the edge of the BEC thus hardly interacting with it. 
On the other hand, the bosonic gas due to its collisions with $\rho^{(1)}_{I}(x;t)$ at the edges of $\rho^{(1)}_{B}(x;t)$ undergoes 
a dipole motion [Fig. \ref{fig:14} (c)]. 

We also remark here that an energy transfer process from the impurities to the bosonic bath occurs in both driving scenarios (results not shown here for brevity). 
Here, the energy gain of the bath is significantly enhanced for $\omega_D=0.6$ where the components are miscible during the evolution 
while for $\omega_D=1.5$ $E_B(t)$ mainly increases for $t<t_f$ since the components overlap and remains almost constant for $t>t_f$ where the 
immiscibility is preserved. 
Additionally, let us note in passing that a similar overall phenomenology regarding the dynamical response of both the bosonic bath and the impurities 
takes place for a continuous shaking of the impurities harmonic oscillator (results not shown for brevity).

\section{Continuous shaking of the trap potential}\label{continuous}

Next, we unravel the emergent nonequilibrium dynamics of the binary bosonic system when considering that the periodic driving of the 
harmonic trap of the impurities is maintained throughout the evolution and not for just two periods of the pulse driving as in 
Sec. \ref{driving_non_cont}. 
The system parameters are the same as in the previous Sec. \ref{driving_non_cont}, namely the bath comprises of $N_B=100$ bosons 
with $g_{BB} = 0.5$ and $N_I=2$ impurities where $g_{II} = 0.4$. 
The impurity-BEC interaction is $g_{BI} = 0.2$ and thus the species are initially ($t=0$) miscible. 
Both species are trapped in a harmonic oscillator of frequency $\omega = 0.3$ and the system is initialized in its many-body ground 
state. 
Subsequently, from $t=0$ on, the harmonic oscillator of the impurities is periodically driven [see Eq. (\ref{driving_protocol})] for the entire time-evolution. 
The oscillation amplitude is assumed to be $\mathcal{A} = 20\gg R_{TF}\approx8.3$.
\begin{figure*}[ht]
\centering
\includegraphics[width=0.95\textwidth]{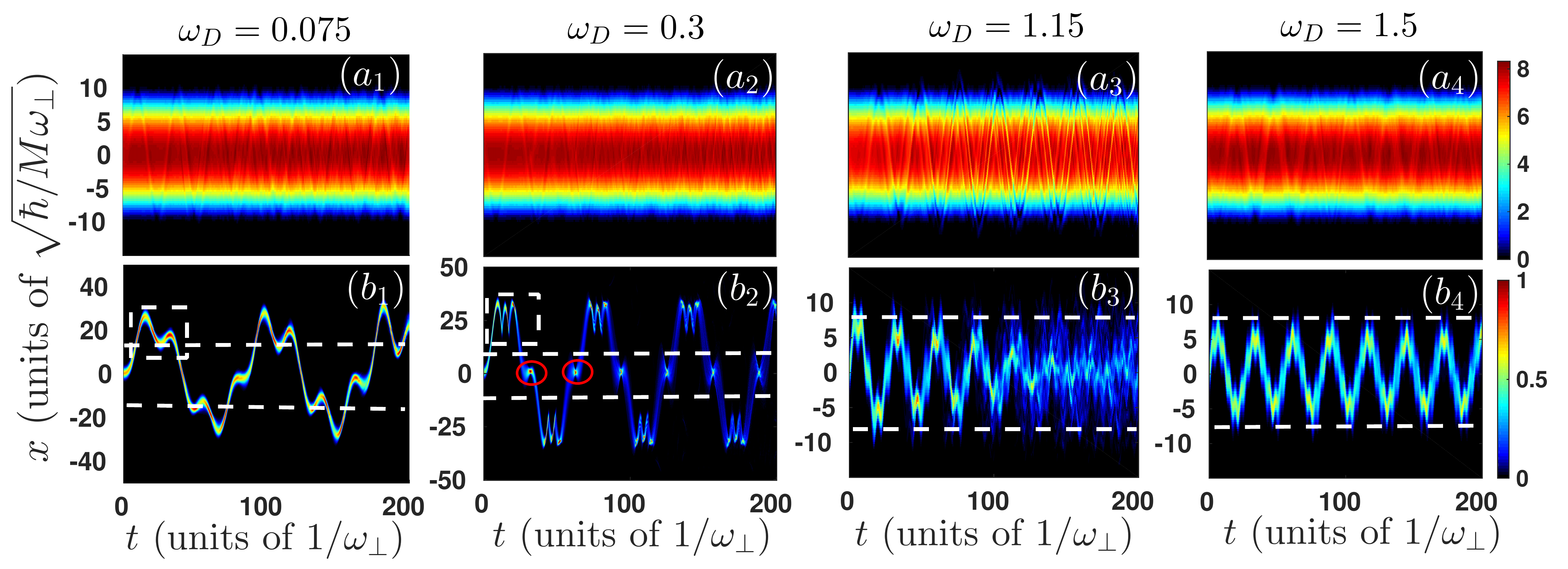}
\caption{Dynamics of the $\sigma$-species single-particle density $\rho^{(1)}_{\sigma}(x;t)$ of ($a_1$)-($a_4$) the bosonic bath ($\sigma = B$) 
and ($b_1$)-($b_4$) the two impurities ($\sigma = I$) for selected shaking frequencies $\omega_D$ (see legends). 
The shaking of the harmonic oscillator potential of the impurities is maintained throughout the evolution. 
The system contains $N_B=100$ bosons and $N_I=2$ impurities with intra- and interspecies interactions 
$g_{BB}=0.5$, $g_{II}=0.4$ and $g_{BI}=0.2$ respectively. 
It is confined in a harmonic trap of frequency $\omega=0.3$ and it is initialized into its ground state. 
The dashed horizontal lines in ($b_1$)-($b_4$) indicate the location of the Thomas-Fermi radius of the bosonic bath. 
The white dashed rectangles in ($b_1$), ($b_2$) and the red circles in ($b_2$) mark specific back-scattering events of the impurities during the driving.}
\label{fig:8}
\end{figure*}

\subsection{Dynamics of the single-particle density and the center-of-mass}\label{density_cont} 

To visualize the nonequilibrium dynamics of the system subjected to a continuous shaking we resort to the time-evolution of the single-particle 
density and the position of the center-of-mass of the $\sigma$-species presented in Fig. \ref{fig:8} and Fig. \ref{fig:10} respectively 
for selective driving frequencies as in Sec. \ref{driving_non_cont}. 
Focusing on low driving frequencies e.g. $\omega_D = 0.075$ the impurities perform an ``irregular'' oscillatory motion overall 
following their driven potential [Fig. \ref{fig:8} ($b_1$)]. 
Note that the observed dynamical response of the impurities is reminiscent of the corresponding response of the case of pulse driving for $t<t_f$ (i.e. before its termination) 
discussed in Sec. \ref{driving_non_cont}, see also Fig. \ref{fig:1} ($b_1$). 
Importantly, here, the motion taking place during the first oscillation period of the external potential is periodically repeated within each 
driving cycle [Figs. \ref{fig:8} ($b_1$)]. 
This behavior of the impurities is also imprinted in their trajectory presented in Fig. \ref{fig:10} ($b_1$). 
We remark that this time periodic behavior of $\rho^{(1)}_{I}(x;t)$ is caused by the continuous driving and it is in contrast to the pulse driven 
case where after the termination of the driving the impurities oscillate well inside the bosonic medium, see also Fig. \ref{fig:1} ($b_1$). 
Also note here that due to the continuous driving the oscillation amplitude of $\rho^{(1)}_{I}(x;t)$ is slightly increased every driving period, e.g. 
$\rho^{(1)}_{I}(x,t=99)$ reaches $x = 29.5$ while it is located at $x = 33.4$ when $t = 182.5$ [Fig. \ref{fig:8} ($b_1$)]. 
Moreover, the BEC background $\rho^{(1)}_{B}(x;t)$ shows weak distortions from its initial Thomas-Fermi profile being imprinted as a small amplitude 
collective dipole motion [Fig. \ref{fig:8} ($a_1$)] during the dynamics. 
This latter motion is directly captured by the time-evolution of the center-of-mass which exhibits multifrequency oscillations as shown in Fig. \ref{fig:10} ($a_1$). 

For resonant driving frequencies, i.e. $\omega_D =\omega= 0.3$, the impurities exhibit an oscillatory behavior moving inside and outside the 
bosonic bath [Fig. \ref{fig:8} ($b_2$)] during the dynamics. 
{Initially they move towards the left edge of the BEC, escaping from the latter at $t\approx 20.2$, and reach $x\approx 33.6$ where they experience 
during their motion two back-scattering events due to the external driving [see the white dashed rectangle in Fig. \ref{fig:8} ($b_2$)]. 
Later on they penetrate the BEC background interacting with its atoms and featuring a dramatic back-scattering event at $x\approx 0$ manifested by 
the prominent density hump of $\rho^{(1)}_{I}(x; t)$ [see the red circle in Fig. \ref{fig:8} ($b_2$)]. 
This behavior suggests that the impurities slow down at this location and bunch momentarily before moving to the opposite edge of the bath. 
Afterwards the impurities undergo a similar to the above-described dynamical behavior, i.e. moving outside the right edge of the BEC and being reflected backwards, 
until the first driving period is completed. 
As time evolves the impurities repeat the same pattern within every driving period, see Fig. \ref{fig:8} ($b_2$). 
Indeed, inspecting the trajectory of the impurities shown in Fig. \ref{fig:10} ($b_2$) we can directly infer their oscillatory behavior characterized by 
a slightly decaying amplitude. 
The latter signals the dissipation of energy into the bosonic bath as we shall argue in the following section. 
We remark that compared to the pulse driving case the dynamical response of the impurities remains the same for $t<t_f$ but it is significantly 
altered for $t>t_f$ where the driving is terminated and the impurities are mainly deposited outside the edge of the bath throughout the evolution [Figs. \ref{fig:1} ($b_1$)]. 
On the other hand the shape of the bosonic single-particle density is to a large extent unperturbed, see Fig. \ref{fig:8} ($a_2$), because the impurities 
reside majorly outside of $\rho^{(1)}_{I}(x;t)$. 
However, the disturbances caused by the motion of the impurities within the bosonic bath are imprinted into the latter as a collective dipole mode identified 
by the oscillatory behavior of $\langle X_{B} (t)\rangle$ depicted in Fig. \ref{fig:10} ($a_2$). 
Interestingly the oscillation amplitude of $\langle X_{B} (t)\rangle$ is amplified over time, a behavior that stems from the continuous nature of the 
driving \cite{mistakidis2015resonant,mistakidis2017mode,Goldman2015}. 
\begin{figure}[ht]
	\centering
	\includegraphics[width=0.46\textwidth]{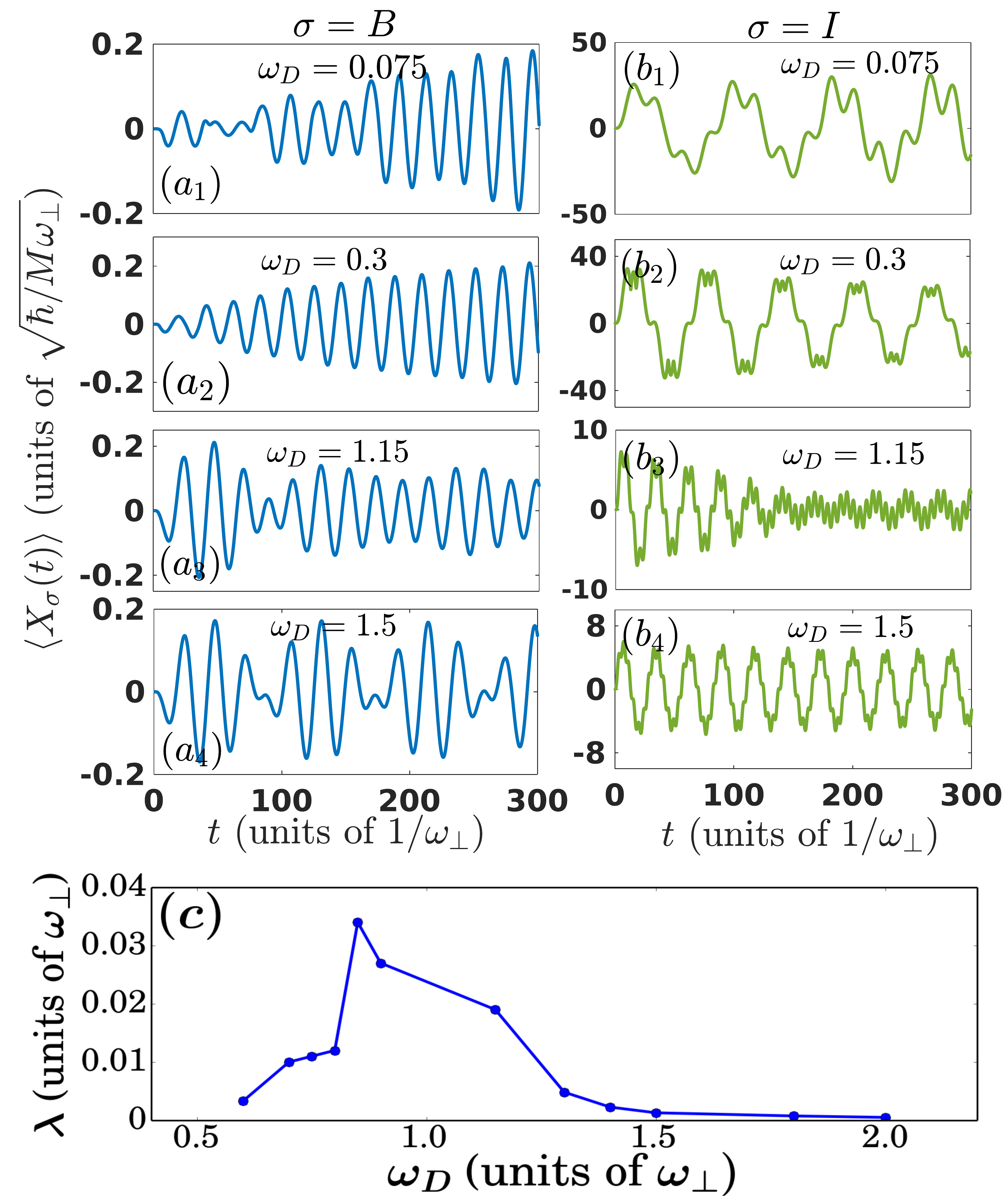}
	\caption{Evolution of the center-of-mass of ($a_1$)-($a_4$) the bath $\langle X_{B}(t) \rangle$ and ($b_1$)-($b_4$) the impurities $\langle X_{I}(t) \rangle$ 
	at distinct driving frequencies $\omega_D$ (see legends). 
	($c$) Damping rate, $\lambda$, of $\langle X_{I}(t) \rangle$ with varying driving frequency for $\omega_D>0.6$. 
	The shaking of the impurities is maintained for the entire evolution. 
	The Thomas-Fermi radius of the bosonic gas is $R_{TF}\approx 8.3$.
	Other system parameters are the same as in Fig. \ref{fig:8}. }
	\label{fig:10}
\end{figure} 

Consequently we inspect the impurities dynamics for large driving frequencies, namely $\omega_D\gg \omega$. 
Here due to the fast shaking of the harmonic oscillator it is very difficult for the impurities to instantaneously follow their external potential and as a result 
their motion is restricted to a spatial region which is smaller than the actual driving amplitude $\mathcal{A}$. 
The resulting time-evolution of $\rho^{(1)}_{I}(x,t)$ is presented in Figs. \ref{fig:8} ($b_3$) and ($b_4$) for driving frequencies $\omega_D=1.15$ and $\omega_D=1.5$ respectively. 
As it can be seen, $\rho^{(1)}_{I}(x;t)$ performs in both cases a decaying amplitude oscillatory motion within the bosonic bath throughout the dynamics. 
This dynamical response of the impurities is also evident in the time-evolution of their trajectory shown in Figs. \ref{fig:10} ($b_3$), ($b_4$). 
Additionally we can deduce that the decay of the oscillation amplitude of the impurities is more pronounced for $\omega_D=1.15$ than $\omega_D=1.5$, a behavior 
that is clearly captured in both the dynamics of $\rho^{(1)}_{I}(x;t)$ [Figs. \ref{fig:8} ($b_3$), ($b_4$)] and the impurities trajectory [Figs. \ref{fig:10} ($b_3$), ($b_4$)]. 
We remark that the larger decay amplitude e.g. of $\braket{X_I(t)}$ for $\omega_D=1.15$ compared to $\omega_D=1.5$ is caused by the enhanced degree 
of interspecies correlations in the former case leading to a faster dephasing of the underlying many-body state, see also Fig. \ref{fig:12} ($a$) and the discussion below. 
It is also worth mentioning that for $\omega_D=1.15$ and referring to the initial stages of the dynamics $\rho^{(1)}_{I}(x;t)$ possesses a localized distribution. 
However deeper in the evolution, $t>100$ in Fig. \ref{fig:8} ($b_3$), where the impurities feature multiple collisions with the atoms of the bosonic gas, 
$\rho^{(1)}_{I}(x;t)$ exhibits a rather delocalized shape. 
Interestingly for even larger driving frequencies, e.g. $\omega_D=1.5$, $\rho^{(1)}_{I}(x;t)$ exhibits a relatively localized configuration. 
We remark that compared to the pulse driving scenario [Figs. \ref{fig:1} ($b_3$), ($b_4$)] the impurities exposed to a continuous shaking remain to a larger extent 
trapped inside their host during the evolution and the decay of their oscillation amplitude is more pronounced, see e.g. Fig. \ref{fig:3} ($b_3$) and Fig. \ref{fig:10} ($b_3$). 
Moreover as a result of the motion of the impurities within the bosonic gas density dips build upon $\rho^{(1)}_{B}(x;t)$ [Figs. \ref{fig:8} ($a_3$), ($a_4$)] at the instantaneous 
location of the density humps of $\rho^{(1)}_{I}(x;t)$. 
Notice that these density dips of $\rho^{(1)}_{B}(x;t)$ become very shallow for $\omega_D=1.5$. 
Overall, the bosonic medium undergoes a multifrequency dipole motion which is captured by its center-of-mass motion presented in Figs. \ref{fig:10} ($a_3$), ($a_4$). 
Also, here the amplitude of this dipole motion seems quite insensitive to the driving frequency, compare Figs. \ref{fig:10} ($a_3$) and ($a_4$).

We have identified that the impurities subjected to a continuous shaking of their harmonic trap remain completely trapped in their host only for large 
driving frequencies, and in particular for $\omega_D>0.6$. 
As a result, we are able to model the decaying motion of the impurities inside the bosonic medium according to the well-known effective damped equation of motion 
\begin{equation}\label{1}
\ddot{x} + \lambda \dot{x} + \omega^2_{\rm eff}x = F_{0}\sin(\omega_D t). 
\end{equation} 
In this expression, $\lambda$ is the damping parameter of the impurities, $\omega_{eff}$ denotes the effective trapping due to the presence of 
the bath and the external harmonic confinement. 
$F_{0}=\mathcal{A}\omega_{eff}^2$ is the amplitude of the external driving force. 
Moreover, by solving Eq.~\ref{1} it can be easily shown that the mean position of the impurities reads
\begin{equation}\label{2}
\begin{split}
\braket{X_{I}(t)} &= e^{-\frac{\lambda}{2}t}\bigg [x_{0} \cos(\omega_0 t) + \frac{u_0+ \frac{\lambda}{2} x_{0}}{\omega_0 } \\& \times \sin(\omega_{0} t)   \bigg] 
+ \frac{F_{0}\sin(\omega_D t + \delta)}{(\omega^2_{\rm eff} - \omega^2_D)^2 + \omega_D^2 \lambda^2},
\end{split}
\end{equation}
where $\omega_0 = \sqrt{(\omega_{\rm eff})^2 - \big(\frac{\lambda}{2}\big)^2}$, $x_0\equiv \braket{\Psi(0)|\hat{x}|\Psi(0)}$, $u_0=\mathcal{A} \omega_D$ 
and $\delta$ is a phase factor. 
Evidently, in this equation the unknown parameters are $\lambda$, $\omega_{\rm eff}$ and $\delta$. 
In order to determine these parameters we perform a fitting of the analytical form of $\braket{X_I(t)}$ provided by Eq.~$\ref{2}$ 
with the numerically obtained result of $\braket{X_I(t)}$. 
Figure \ref{fig:10} ($c$) shows the value of the damping term $\lambda$ obtained through the above-described fitting procedure with respect to the 
driving frequency. 
We observe that $\lambda$ increases within the interval $\omega_D\in \{0.6, 0.85\}$ and subsequently shows an overall decreasing tedency. 
This behavior of $\lambda$ is also in line with the growth of the average degree of interspecies correlations captured by $\bar{S}_{VN}\equiv(1/T)\int_0^T dt S_{VN}(t)$. 
The latter increases for $\omega_D\in \{0.6, 0.85\}$ and afterwards decreases, see also Fig. \ref{fig:12} ($a$). 
Accordingly, for $\omega_D>1.5$ where $\bar{S}_{VN}\to0$ also $\lambda\to 0$.

\subsection{Energy exchange processes} \label{energy_cont} 

In order to infer whether impurity-BEC energy transfer mechanisms \cite{lampo2017bose,mistakidis2019dissipative,mistakidis2019quench} occur in the course of 
the continuously driven dynamics we inspect the underlying 
intra- and interspecies energy terms, namely the energy of the bath $E_B(t)$ and the impurities $E_I(t)$ as well as the interspecies interaction 
energy $E_{BI}(t)$ introduced in Sec. \ref{energy_non_cont}. 
Figure \ref{fig:11} illustrates the dynamics of the above-mentioned energy contributions for different driving frequencies. 
We can deduce that independently of $\omega_D$ the energy of the impurities $E_I(t)$ shows an oscillatory behavior [Figs. \ref{fig:11} ($a_1$)-($a_4$)] whilst the energy of 
the bosonic bath $E_B(t)$ overall increases [Figs. \ref{fig:11} ($c_1$)-($c_4$)]. 
More precisely, for a larger $\omega_D$ the oscillatory pattern of $E_I(t)$ involves a larger number of frequencies [see Fig. \ref{fig:11} ($a_1$) and Fig. \ref{fig:11} ($a_4$)] and 
the increase of $E_{B}(t)$ becomes more enhanced, e.g. compare Fig. \ref{fig:11} ($c_1$) with Fig. \ref{fig:11} ($c_4$). 
We remark that the oscillations of $E_{I}(t)$ essentially reflect the impurities motion and in particular when they move to the edge of the BEC they possess a larger kinetic energy 
than if they are close to the trap center, resulting in an increasing tendency of $E_{I}(t)$. 
Notice also that $E_{I}(t)$ maximizes for $\omega_D=\omega$ \cite{Goldman2014,Goldman2015,mistakidis2015resonant} which explains the fact that the impurities 
exhibit the larger oscillation amplitude [Fig. \ref{fig:10} ($b_2$)]. 
\begin{figure}[ht]
\centering
\includegraphics[width=0.48\textwidth]{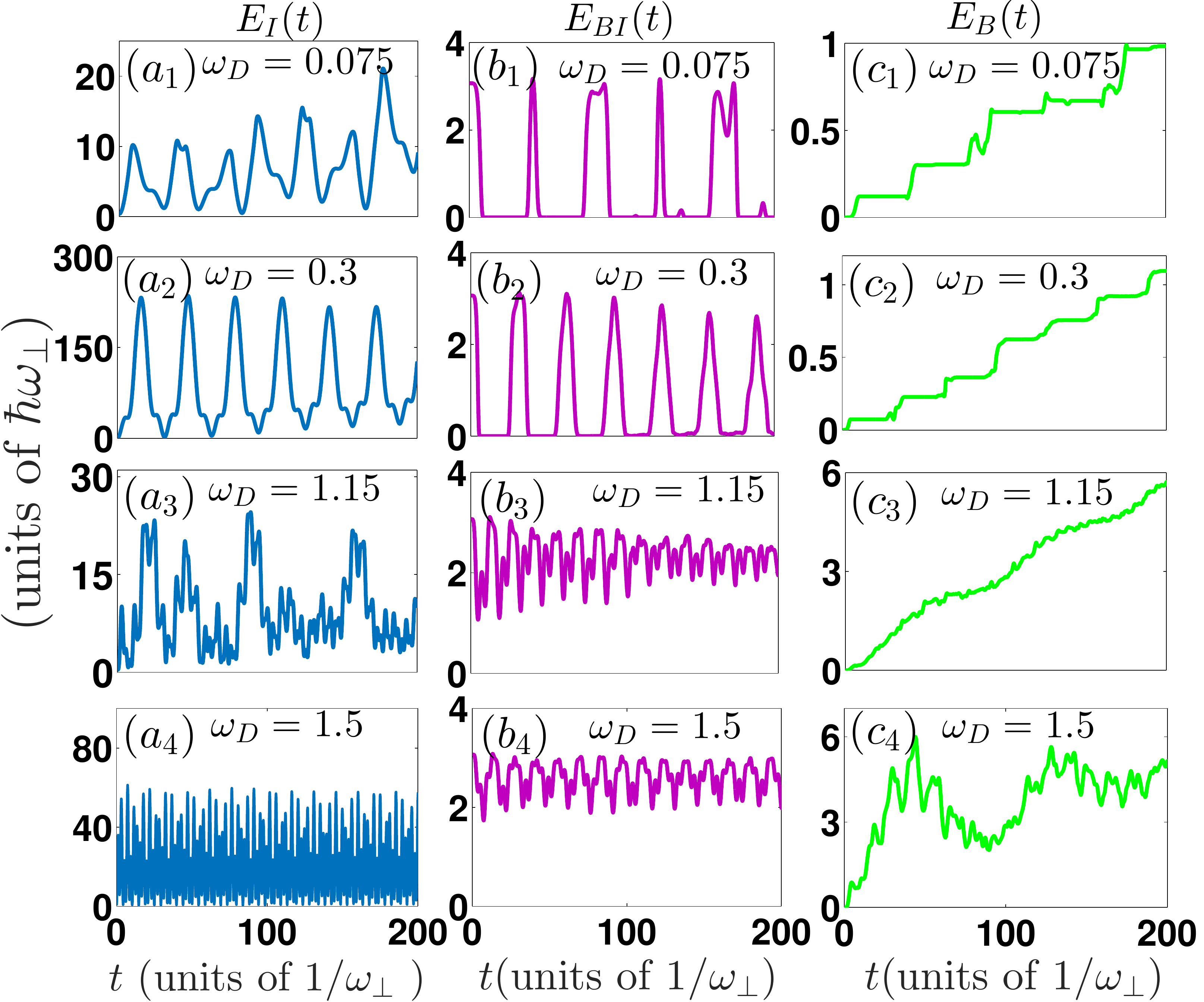}
\caption{Time-evolution of ($a_1$)-($a_4$) the energy of the impurities $E_I(t)$, ($b_1$)-($b_4$) the interspecies interaction energy $E_{BI}(t)$ and 
$(c_1)$-($c_2$) the energy of the bath $E_{B}(t)$ at distinct driving frequencies $\omega_D$ (see legends). 
The shaking of the harmonic oscillator of the impurities is maintained throughout the dynamics. 
The remaining system parameters are the same as in Fig. \ref{fig:8}.}
\label{fig:11}
\end{figure} 

\begin{figure}[ht]
\centering
\includegraphics[width=0.45\textwidth]{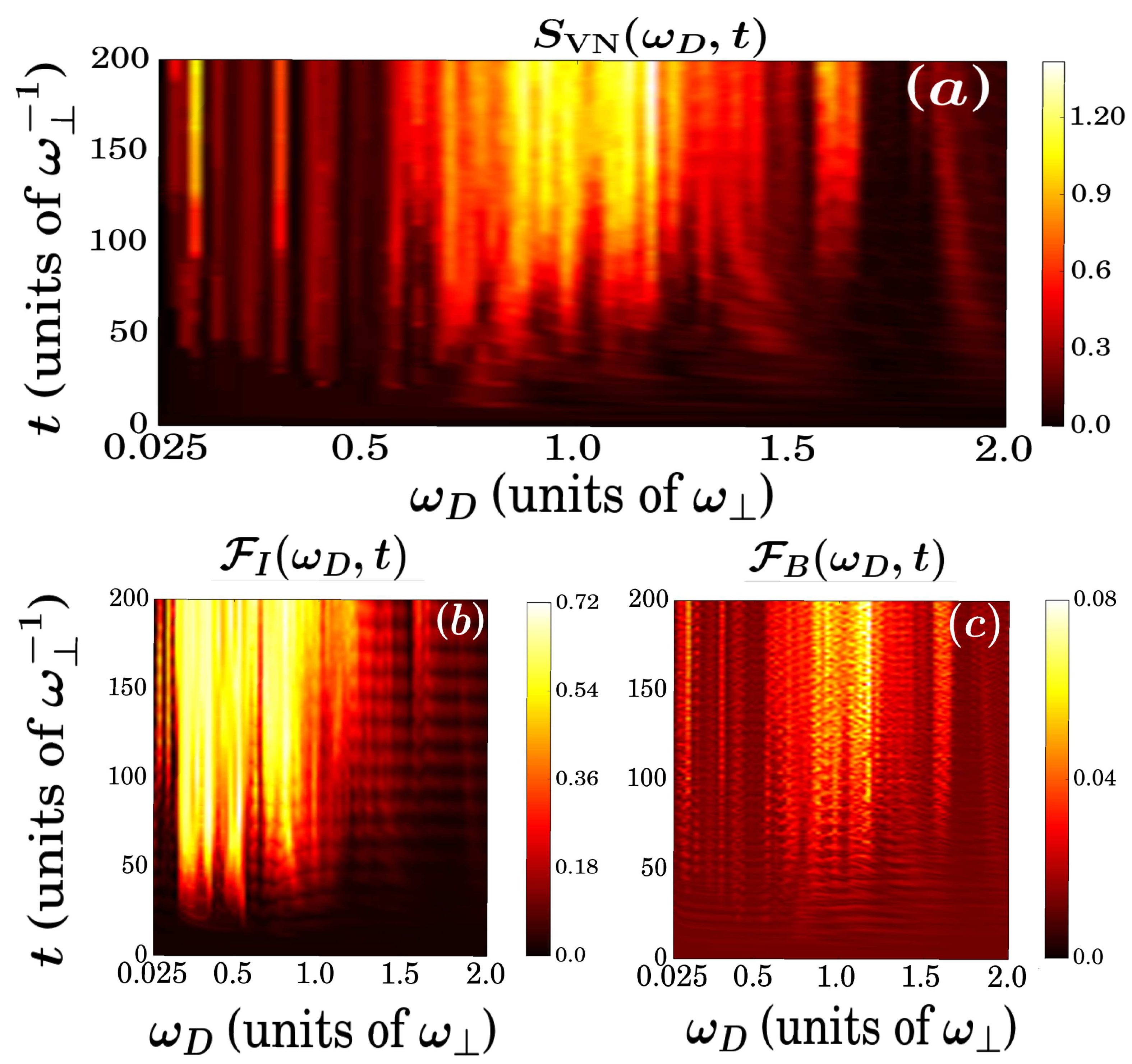}
\caption{Temporal-evolution of (a) the von-Neumann entropy $S_{VN}(t)$ and the deviation from unity of the first natural population of 
(b) the impurities $\mathcal{F}_{I}(t)$ and (c) the bosonic gas $\mathcal{F}_{B}(t)$ as a function of the driving frequency $\omega_D$. 
The harmonic oscillator of the two impurities is subjected to a continuous shaking. 
The remaining system parameters are the same as in Fig. \ref{fig:8}. }
\label{fig:12}
\end{figure}

Most importantly, the simultaneous enhancement of $E_{B}(t)$ accompanied by a reduction of $E_{BI}(t)$ [Figs. \ref{fig:11} ($b_1$)-($b_4$)] indicates the dissipation of energy 
from the impurity to its BEC background \cite{mistakidis2019correlated,mistakidis2019dissipative,nielsen2019critical} manifested in the density of the latter as a collective dipole motion. 
Additionally it is worth commenting that $E_{BI}(t)$ shows a significantly distinct behavior for low and high driving frequencies. 
For instance, in the former case $E_{BI}(t)$ becomes zero at specific time-intervals [Fig. \ref{fig:11} ($b_2$)] where the impurities escape from the bosonic 
gas [Fig. \ref{fig:8} ($b_2$)] and as a consequence $E_B(t)$ shows a constant plateau in the same time frame [Figs. \ref{fig:11} ($c_1$), ($c_2$)]. 
However for fast drivings $E_{BI}(t)$ performs irregular oscillations while remaining finite throughout the evolution [Fig. \ref{fig:11} ($b_4$)] 
since the impurities reside well inside the BEC background [Fig. \ref{fig:8} ($b_4$)].

\subsection{Intra and interspecies correlations} 

To testify the importance of beyond mean-field intra- and interspecies correlations during the evolution of the system we next employ $\mathcal{F}_{\sigma}(t)$ 
[Eq. (\ref{fragmentation_measure})] and $S_{VN}(t)$ [Eq. (\ref{entropy})] respectively. 
Recall that $\mathcal{F}_{\sigma}(t)>0$ indicates the existence of $\sigma$-species intraspecies correlations and $S_{VN}(t)\neq0$ 
designates the occurrence of interspecies ones \cite{mistakidis2018correlation,mistakidis2019correlated}. 
Figure \ref{fig:12} showcases both $S_{VN}(t)$ and $\mathcal{F}_{\sigma}(t)$ for a wide range of $\omega_D$. 
Evidently, for short evolution times $t<40$ both intra- and interspecies correlations of the system are suppressed since $\mathcal{F}_{\sigma}(t)$ and $S_{VN}(t)$ 
deviate only slightly from zero, e.g. $\mathcal{F}_{B}(t=20)\approx 0.015$, $\mathcal{F}_{I}(t=20)\approx 0.02$ and $S_{VN}(t=20)\approx 0.04$ at $\omega_D=0.1$. 
However for $t>40$ we observe a significant development of impurity-BEC [Fig. \ref{fig:12} ($a$)] and impurity-impurity [Fig. \ref{fig:12} ($b$)] correlations 
while the intraspecies correlations of the bosonic gas remain adequately small for every $\omega_D$ [Fig. \ref{fig:12} ($c$)]. 
Indeed, the largest value of $\mathcal{F}_{B}(t)$ occurs around $\omega_D=1.25$ where $\mathcal{F}_{B}(t=180)\approx0.07$. 
In particular the impurity-BEC and impurity-impurity correlations, as captured via $S_{VN}(t)$ and $\mathcal{F}_{I}(t)$, are maximized in 
the range $0.6<\omega_D<1.7$ and $0.1<\omega_D<1.5$ respectively. 
The predominantly negligible entanglement for $\omega_D<0.3$ can be attributed to the fact that the impurities majorly lie 
outside of the BEC background in the course of the evolution. 
Note also here that despite the weak entanglement the impurities appear to be strongly correlated for these driving frequencies. 
Furthermore, for $0.6<\omega_D<1.7$ [Fig. \ref{fig:12}($a$)] the impurities are trapped within the bath throughout the evolution [Fig. \ref{fig:8} ($b_4$)], testifying 
the increasing tendency of $S_{VN}(t)$ compared to other values of $\omega_D$. 
At $\omega_D>1.7$ both $S_{VN}(t)$ and $\mathcal{F}_{I}(t)$ acquire very small values, a behavior that is attributed to the high frequency driving where the impurities 
motion cannot be synchronized with the external driving. 
\begin{figure}[ht]
\centering
\includegraphics[width=0.48\textwidth]{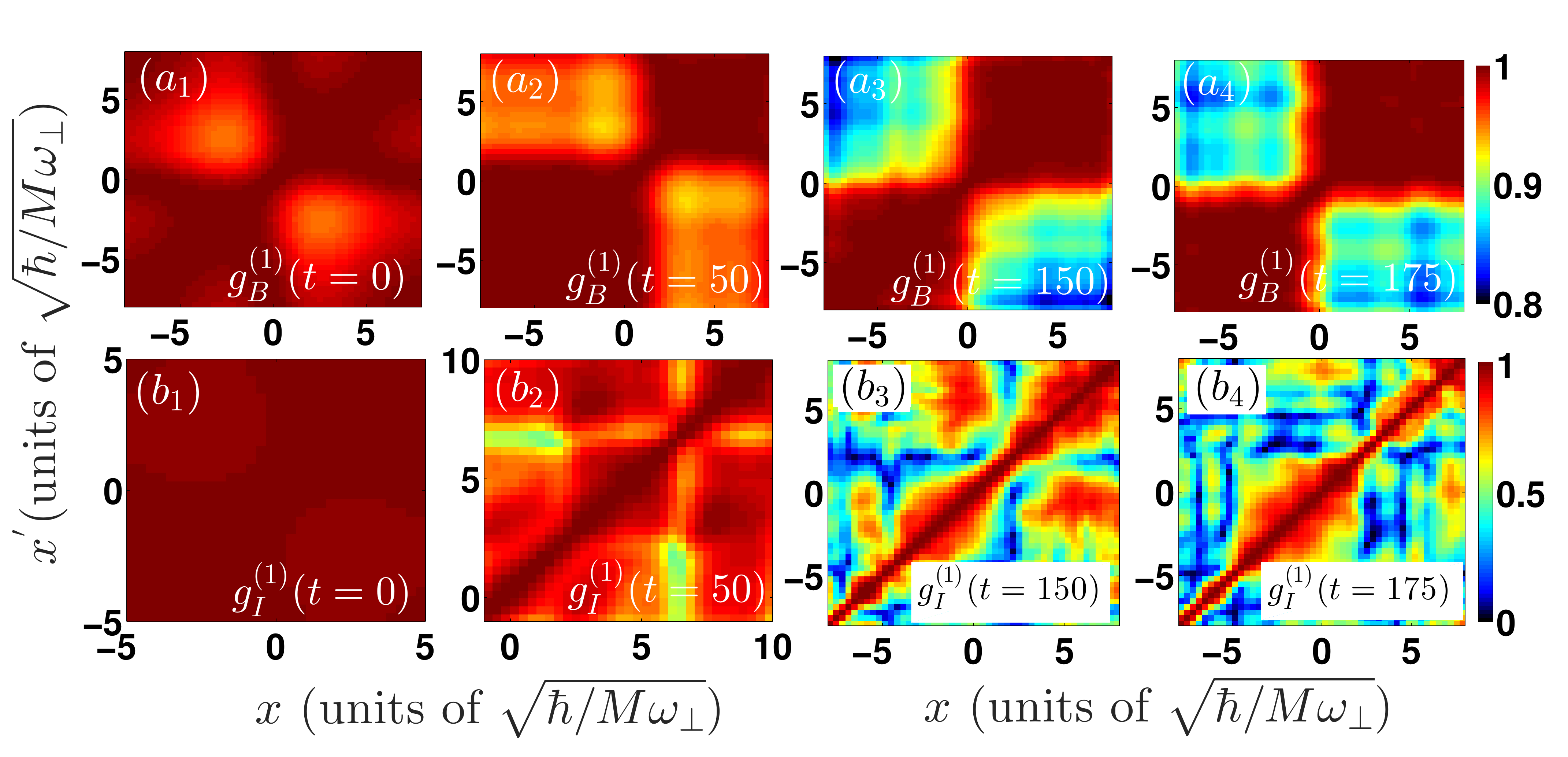}
\caption{One-body coherence function $g^{(1)}_{\sigma}(x,x';t)$ of ($a_1$)-($a_4$) the bosonic bath and ($b_1$)-($b_4$) the impurities 
at specific time-instants of the evolution (see legends). 
The dynamics is induced by a continuous shaking of the harmonic oscillator potential of the impurities at a driving frequency $\omega_D=1.15$. 
The remaining system parameters are the same as in Fig. \ref{fig:8}.}  
\label{fig:13}
\end{figure} 

In view of the above, for $0.6<\omega_D<1.7$ intra- and interspecies correlations are finite testifying the necessity of a beyond mean-field 
treatment of the dynamics. 
For $\omega_D<0.7$ since $S_{VN}(t)$ and $\mathcal{F}_{B}(t)$ are suppressed a corresponding product state on the species and the BEC level constitutes 
an adequate approximation. 
However, the impurities wavefunction is a superposition of the different single-particle states due to $\mathcal{F}_{I}(t)>0$. 
Finally, for $\omega_D>1.7$ all interparticle correlations almost vanish and therefore the dynamics of the mixture can be modeled to a good approximation 
within a corresponding mean-field treatment, i.e. $\lambda_1(t)=1$ in Eq. (\ref{wfn_ansatz}) and $n_1^B(t)=n_1^I=1$ in Eq. (\ref{SPF}).

\subsection{Coherence losses} \label{coherence_cont}

Subsequently, we unravel losses of the $\sigma$-species spatial coherence \cite{Naraschewski1999,mistakidis2018correlation,mistakidis2019correlated,Katsimiga_2017} by invoking 
the corresponding one-body coherence function $g_{\sigma}^{(1)}(x,x';t)$ [Eq. (\ref{one_body_coherence})]. 
As in Sec. \ref{coherence_non_cont} we showcase the case of a large driving frequency, $\omega_D=1.15$, due to the significant role of correlations 
in this driving regime compared to the others, see also Fig. \ref{fig:12}. 
Snapshots of $g^{(1)}_{B}(x, x',t)$ and $g^{(1)}_{I}(x, x';t)$ are shown in Figs. \ref{fig:13} ($a_1$)-($a_4$) and ($b_1$)-($b_4$) respectively. 
Closely inspecting $g^{(1)}_{B}(x, x^{'},t)$ we can infer that only very small coherence losses take place between the spatial regions $0<x<6$ and $-6<x'<0$. 
These losses of coherence are almost negligible for $t<50$, e.g. $g^{(1)}_{B}(x=5, x'=-1,t=50)\approx0.97$ in Fig. \ref{fig:13} ($a_2$), and later on become relatively 
pronounced, see e.g. $g^{(1)}_{B}(x=5, x'=-6,t=175)\approx0.82$ in Fig. \ref{fig:13} ($a_4$). 
Note that this behavior of $g^{(1)}_{B}(x,x',t)$ is in line with the suppressed degree of intraspecies correlations of the bath presented in Fig. \ref{fig:12} ($c$). 
Also, the amount of coherence losses is slightly increased when compared to the pulse driving scenario [Figs. \ref{fig:6} ($a_1$)-($a_4$)]. 
Regarding the impurities we observe that at $t=0$ they are almost perfectly coherent since $g^{(1)}_{I}(x, x';t=0)>0.98$ for every $x$, $x'$. 
However as time evolves a systematic build up of coherence losses \cite{li2019controlling} for $x\neq x'$ occurs, e.g. $g^{(1)}_{I}(x=2,x'=3;t=50) \approx 0.95$ in Fig. \ref{fig:13} ($b_2$), 
which becomes enhanced for longer times, e.g. $g^{(1)}_I(x = 2; x'=-2;t=150)\approx0.18$ [Fig. \ref{fig:13} ($b_3$)] and 
$g^{(1)}_I (x=2.5, x'=-2.5; t=175)\approx0.3$ [Fig. \ref{fig:13} ($b_4$)]. 
As previously, the emergent coherence losses are visualized in $g^{(1)}_I (x; x';t)$ via the suppression of its off-diagonal elements. 
\begin{figure}[ht]
\centering
\includegraphics[width=0.48\textwidth]{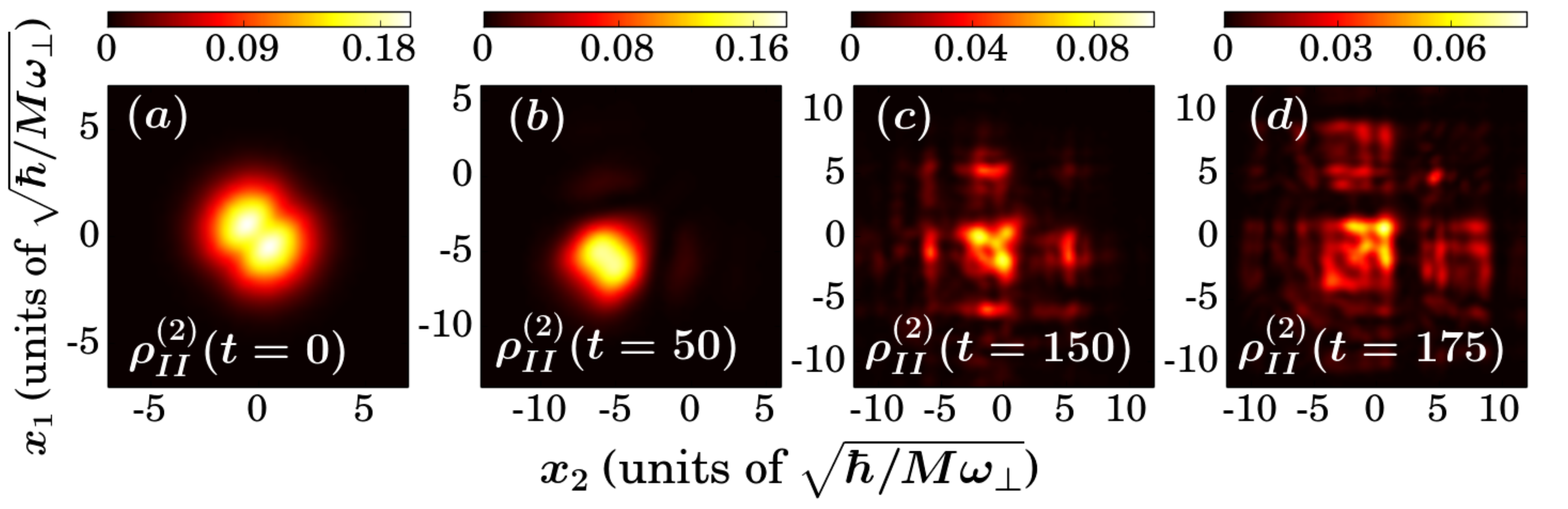}
\caption{Instantaneous two-body reduced density matrix of the impurities at different time-instants of the evolution (see legends) following a 
continuous shaking of the harmonic oscillator at frequency $\omega_D=1.15$. 
The remaining system parameters are the same as in Fig. \ref{fig:8}.}  
\label{fig:18}
\end{figure}

\subsection{Two-body dynamics of the impurities} \label{two_body_reduced_cont} 

Next, we monitor the spatially resolved dynamics of the two impurities with respect to one another by resorting to 
the diagonal of the two-body bosonic reduced density matrix 
\begin{equation}
\begin{split}
\rho^{(2)}_{II}(x_1,x_2;t)=\bra{\Psi_{MB}(t)}\hat{\Psi}^{I \dagger}(x_1) \hat{\Psi}^{I \dagger}(x_2)\\ \times \hat{\Psi}^{I}(x_1) \hat{\Psi}^{I}(x_2)\ket{\Psi_{MB}(t)}. 
\end{split}
\end{equation} 
In this expression, $\hat{\Psi}^{I}(x_1)$ is the corresponding bosonic field operator that annihilates a boson at position $x_1$. 
Recall that $\rho^{(2)}_{II}(x_1,x_2;t)$ provides the probability of measuring simultaneously one boson to be located at $x_1$ and the other 
one at $x_2$ \cite{mistakidis2018correlation,mistakidis2019correlated,Naraschewski1999}. 
For our investigation we focus on large driving frequencies where the impurities reside within the bosonic gas throughout the evolution [see also Figs. \ref{fig:8} ($b_3$), ($b_4$)] 
and also the interparticle correlations of the system are enhanced [Fig. \ref{fig:12}]. 
Moreover since the impurities are trapped within the bosonic gas they are dressed by its excitations forming quasiparticles. 
Consequently, these quasiparticles can either move independently or interact thereby forming a pair 
\cite{theel2019entanglement,dehkharghani2018coalescence,mistakidis2019many,mistakidis2019induced}. 

Figure \ref{fig:18} depicts $\rho^{(2)}_{II}(x_1,x_2;t)$ at certain time-instants of the evolution upon considering a continuous shaking of the 
impurities harmonic oscillator at $\omega_D=1.15$. 
Initially $t=0$ [Fig. \ref{fig:18} (a)] the two bosons reside together at the trap center as $\rho^{(2)}_{II}(-2<x_1<2,-2<x_2<2;t=0)$ exhibits a high two-body 
probability peak in the domain $-2<x_1,x_2<2$. 
As time evolves the impurities oscillate within the bosonic bath [see also Fig. \ref{fig:8} ($b_3$)] as a pair since 
$\rho^{(2)}_{II}(-8<x_1<-2,-8<x_2<-2;t=50)$ is predominantly populated [Fig. \ref{fig:18} ($b$)]. 
Simultaneously signatures of a delocalized behavior are observed due to the small values of the off-diagonal elements of $\rho^{(2)}_{II}(x_1,x_2;t=50)$. 
Entering deeper in the evolution the aforementioned delocalization of the impurities becomes more prominent since $\rho^{(2)}_{II}(x_1,x_2;t)$ disperses 
as shown in Figs. \ref{fig:18} ($c$), ($d$). 
This dispersive behavior of $\rho^{(2)}_{II}(x_1,x_2;t)$ is inherently related to the one exhibited by $\rho^{(1)}_{I}(x,t)$ in Fig. \ref{fig:8} ($c$) suggesting 
from a two-body perspective the involvement of excited states in the impurity dynamics. 
Most importantly, the diagonal of $\rho^{(2)}_{II}(x_1,x_2;t)$ is predominantly populated [Figs. \ref{fig:18} ($c$), ($d$)] 
which is suggestive of the presence of attractive induced impurity-impurity interactions 
\cite{theel2019entanglement,mistakidis2019correlated,dehkharghani2018coalescence,mistakidis2019many,mistakidis2019induced}. 
Similar pairing mechanisms of bosonic impurities mainly concentrating on the stationary properties of bosonic mixtures have 
been discussed in Refs. \cite{dehkharghani2018coalescence,camacho_2018,klein2005interaction}.

\section{Conclusions}\label{conclusion}

We have investigated the driven dynamics of two repulsively interacting impurities immersed in a bosonic bath following two different shaking protocols 
of the harmonic trap of the impurities. 
Namely, the shaking is either performed via a pulse consisting of two driving periods and then the system is left to evolve unperturbed or it is 
maintained throughout the evolution corresponding to a continuous driving. 
A particular focus has been placed on setups where the impurities and the bath are initially spatially overlapping (miscible components) 
while the case of initially immiscible components has also been briefly discussed. 
Moreover the dynamical response of the impurities has been carefully explored for a wide range of driving frequencies ranging from low 
to high frequency driving and has been characterized by utilizing several diagnostics including one- and two-body observables as well as 
the individual energy contributions of the species. 

Regarding the pulse driving scenario and for initially miscible (overlapping) components we have identified different dynamical 
response regimes of the impurities depending on the driving frequency as compared to the frequency of the harmonic trap. 
For low driving frequencies, in the course of the shaking the impurities oscillate in space within and outside their host closely following the motion of their trap. 
However after the termination of the pulse their oscillation amplitude decays and they are trapped in the bosonic gas. 
Entering the resonant driving regime, i.e. for a driving frequency close to the harmonic oscillator one, the impurities undergo a more complex dynamics. 
Namely in the duration of the shaking they perform large amplitude irregular oscillations escaping and re-entering into the bosonic gas while afterwards 
they essentially decouple from the bath. 
For large driving frequencies, much larger than the external trap frequency, it is shown that the impurities remain predominantly trapped within the 
bosonic gas especially after the pulse has terminated and exhibit a dispersive behavior for long evolution times. 
Turning to initially immiscible components we have shown that despite the intricate tendency for zero spatial overlap, the impurities subjected to moderate 
drivings feature a dispersive behavior within the bosonic gas after the pulse is terminated.  
However a vigorous shaking renders the impurities to oscillate around the edges of the Thomas-Fermi background of the bosonic bath, thus preserving 
their spatial separation with the bath almost intact. 

Considering a continuous shaking of the trap of the impurities for miscible components we observed that a similar overall phenomenology as for the pulse driven case 
takes place especially for very low and high driving frequencies. 
However, the dynamical response of the impurities here is periodically repeated in time due to the driving protocol and regarding the high frequency driving the impurities 
are found to be better trapped into their host compared to the pulse driving. 
Most importantly, it is showcased that in the resonantly driven regime the impurities perform a periodic oscillatory motion, moving within and escaping from the BEC background, 
while featuring multiple collision events with the latter. 
Furthermore independently of the driving frequency and the protocol the motion of the impurities perturbs the bosonic bath which is subsequently excited performing 
a collective dipole motion. 
Also these excitations are more prominent for high frequency drivings where the impurities mostly reside within their host. 

Examining the individual energy contributions of each species we reveal that when the impurities are trapped into the bosonic bath they transfer energy to the latter, 
a behavior that is more pronounced for large driving frequencies. 
We expose the participation of inter- and intraspecies correlations during the dynamics and show that their degree is enhanced for high driving frequencies. 
The development of coherence losses both in the bosonic gas and the impurities is unveiled and most importantly it is found that the impurities predominantly move 
as a pair and not individually. 

There are several promising research directions, based on the present work, to be considered in future endeavors. 
A straightforward one is to employ two fermionic impurities immersed either in a bosonic or a fermionic environment 
and investigate the emergent periodically driven dynamics induced by the protocol used herein. 
Another interesting perspective is to construct in the continuous low frequency driving case an effective model 
according to which the impurities are dressed by the excitations of the BEC when they lie inside the latter but they are undressed during the time-intervals 
that they reside outside their host. 
Within such a model it might be possible to identify the corresponding polaronic properties of the impurities such as their effective mass and 
induced interactions \cite{mistakidis2019induced,mistakidis2019effective}. 
Moreover, the driven dynamics of impurities trapped in an optical lattice \cite{Keiler2018} instead of a harmonic trap in order to 
control their transport properties is an interesting perspective. 
Finally, in the framework of the present work, the simulation of the corresponding contrast by considering spinor impurities in order to identify 
possibly emerging polaronic states \cite{Mistakidis2019} is definitely worth pursuing.

%%%%%%%%%%%%%%%%%%%%%%%%%%%%%%%%%%%%%%%%%%%%%%%%%%%%%%%%%%%%%%%%%%%%%%%%%%%%%
%%%%%%%%%%%%%%%%%%             Acknowledgements             %%%%%%%%%%%%%%%%%
%%%%%%%%%%%%%%%%%%%%%%%%%%%%%%%%%%%%%%%%%%%%%%%%%%%%%%%%%%%%%%%%%%%%%%%%%%%%% 

\begin{acknowledgments}
K.M. acknowledges a research fellowship (Funding ID no 57381333)  from  the Deutscher Akademischer Austauschdienst (DAAD). 
S.I.M and P.S. acknowledge financial support by the Deutsche Forschungsgemeinschaft (DFG) in the framework of the SFB 925 Light induced 
dynamics and control of correlated quantum systems. 
S. I. M  gratefully acknowledges financial support in the framework of the Lenz-Ising Award of the University of Hamburg.
\end{acknowledgments}
%\clearpage

\appendix

\section{Periodically driven dynamics of a mass-imbalanced mixture}\label{imbalanced_mixture}

In the main text, all of the presented results have been focusing on mass balanced mixtures. 
Another interesting scenario is to consider a mass-imbalanced system and in particular the case of heavy impurities 
immersed in the bosonic bath in order to inspect whether the mass-imbalance can potentially alter the nonequilibrium dynamics 
discussed in Sec. \ref{continuous}. 
Such a typical mass-imbalance bosonic mixture corresponds to a $^{87}$Rb bath and two $^{133}$Cs impurities prepared at the hyperfine states 
$\Ket{F=1, m_F=0}$ and $\Ket{F=3, m_F=2}$ respectively and trapped in the same harmonic oscillator \cite{hohmann2015neutral}. 
\begin{figure}[ht]
\centering
\includegraphics[width=0.45\textwidth]{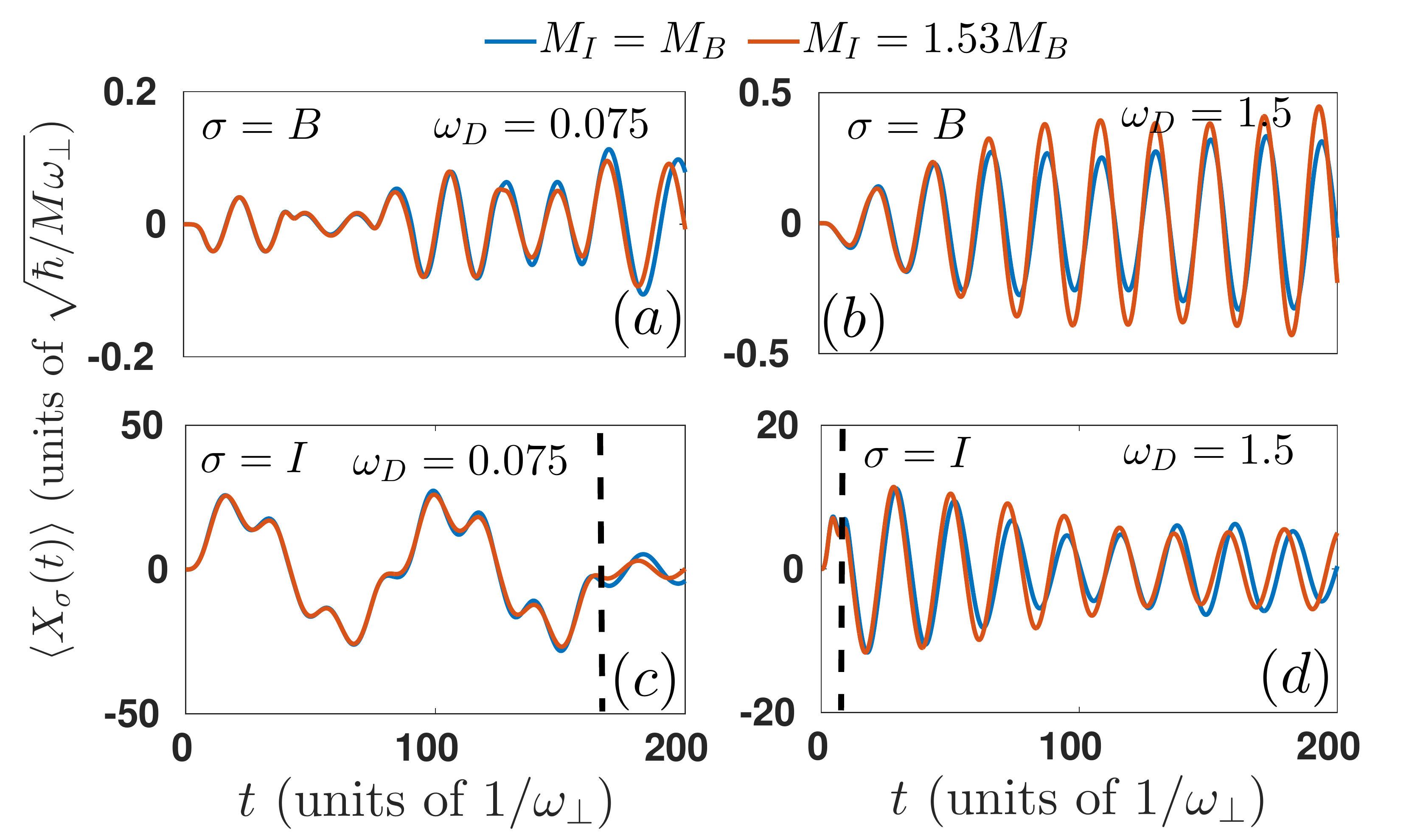}
\caption{Comparison of the temporal-evolution of the center-of-mass of the bosonic bath $\langle X_{B}(t) \rangle$ (upper panels) and the impurities $\langle X_{I}(t) \rangle$ (lower panels) 
between a mass-balanced and a mass-imbalanced mixture (see legend) at different driving frequencies $\omega_D$ (see legend). 
The pulse shaking of the harmonic trap of the impurities is performed for two driving periods until $t=t_f=4 \pi /\omega_D$ (see the vertical lines) and 
afterwards the system is left to evolve unperturbed. 
The remaining system parameters are the same as in Fig. \ref{fig:1}.} 
\label{fig:17}
\end{figure} 

The time-evolution of the center-of-mass oscillation of the impurities $\langle X_{I}(t) \rangle$ and the bosonic gas $\langle X_{B}(t) \rangle$, following a pulse 
shaking of the harmonic trap of the impurities, is shown in Fig. \ref{fig:17} for selective driving frequencies and for both a mass-balanced and a mass-imbalanced mixture. 
Overall, we observe that the dynamical response of both the bosonic gas and the impurities is not significantly affected by the mass-imbalance. 
More specifically, for low driving frequencies $\omega_D=0.075$ both $\langle X_{B}(t) \rangle$ [Fig. \ref{fig:17} (a)] and $\langle X_{I}(t) \rangle$ [Fig. \ref{fig:17} (c)] 
are seen to be essentially insensitive to the considered mass ratio. 
A similar behavior is encountered for high driving frequencies e.g. $\omega_D=1.5$ but here the oscillation amplitude of $\langle X_{B}(t) \rangle$ is slightly larger 
in the mass-imbalanced case [Fig. \ref{fig:17} (b)]. 
This is an expected behavior because the heavy impurities can perturb their host to a larger extent compared to the lighter ones due to $m_{Cs}>m_{Rb}$. 
Moreover, tiny deviations occur also in the oscillation amplitude of $\langle X_{I}(t) \rangle$ [Fig. \ref{fig:17} (d)] between the 
mass balanced and imbalanced cases but with no major tendency.

\section{Remarks on the many-body computational methodology}\label{convergence}

As we discussed in Sec. \ref{wavefunction_ansatz}, in order to study the periodically driven nonequilibrium dynamics of the bosonic mixture, we rely on 
the multi-layer multi-configurational time-dependent Hartree method for atomic mixtures (ML-MCTDHX) \cite{cao2017unified,cao2013multi,kronke2013non}. 
It is an \textit{ab-initio} approach for solving the time-dependent Schr{\"o}dinger equation of multicomponent systems with bosonic 
\cite{mistakidis2018correlation, Katsimiga2017,Katsimiga_2017,katsimiga2018many} 
or fermionic ~\cite{Koutentakis2019,Siegl2018} constituents including also spin degrees of freedom \cite{mistakidis2019quench,Koutentakis2019}. 
The main facet of this numerical approach is that the many-body wavefunction is expanded with respect to a time-dependent and variationally optimized basis. 
The latter enables us to span the relevant subspace of the Hilbert space at each time-instant of the dynamics in a more efficient manner when compared to methods employing a time-independent basis. 
Furthermore, its multi-layer ansatz for the total wavefunction is tailored to capture both the intra- and interspecies correlations emerging during the nonequilibrium 
dynamics of a multicomponent system. 

\begin{figure}[ht]
\centering
\includegraphics[width=0.4\textwidth]{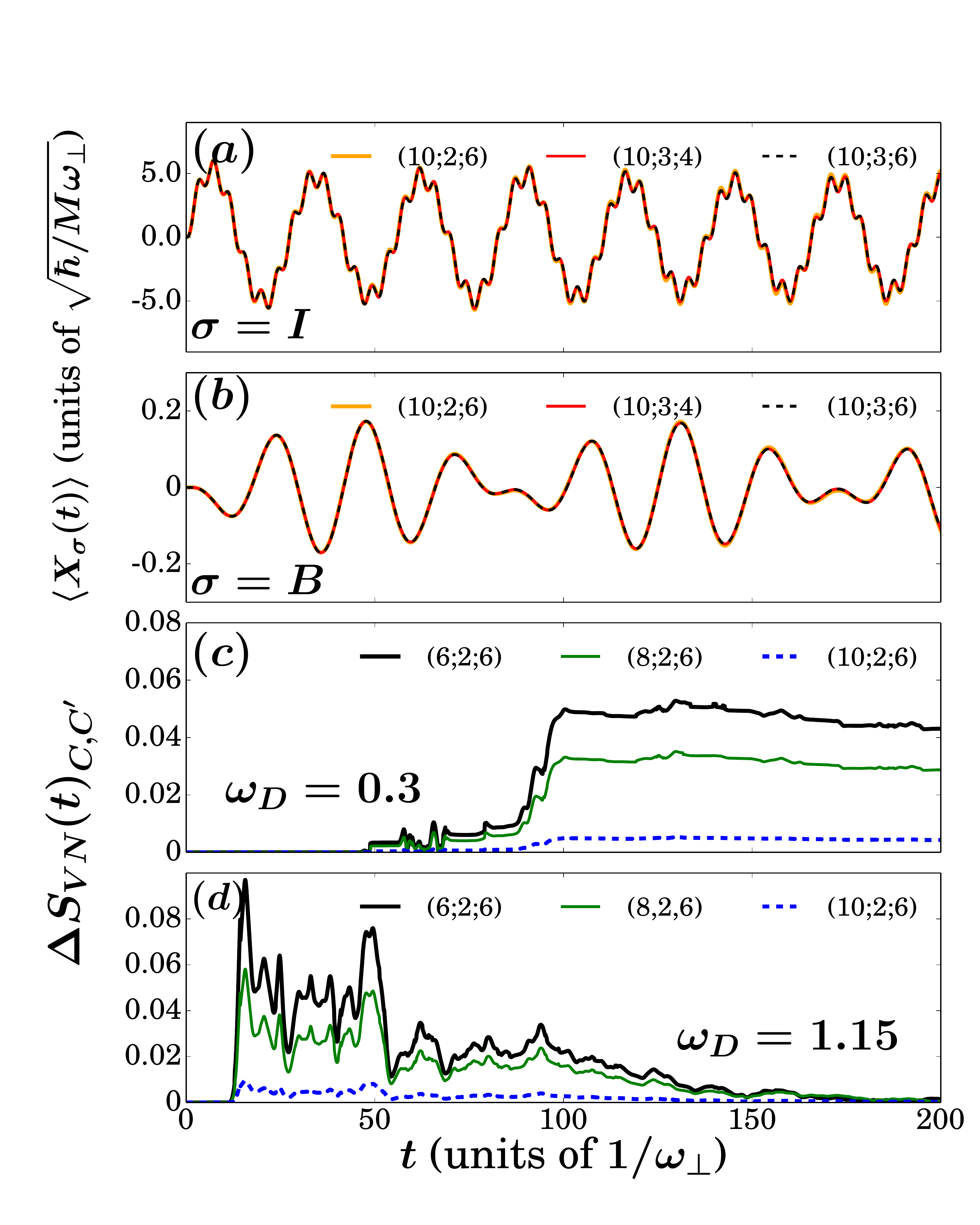}
\caption{Time-evolution of the center-of-mass of ($a$) the two interacting impurities $\langle X_{I}(t) \rangle$ and ($b$) the bosonic gas $\langle X_{B}(t) \rangle$ 
for different orbital configurations ($D;d_{B};d_{I}$) (see legend) at a high driving frequency $\omega_D=1.5$. 
Dynamics of the deviation of the von-Neumann entropy between the $C=(10;3;6)$ and other orbital combinations ($D;d_{B};d_{I}$) (see legend) for driving 
frequency ($c$) $\omega_D=0.3$ and ($d$) $\omega_D=1.15$. 
The shaking of the harmonic oscillator of the impurities is performed throughout the evolution. 
The mixture consists of $N_B=100$ bosons with $g_{BB}=0.5$ and $N_I=2$ interacting $g_{II}=0.4$ impurities in a harmonic trap of frequency $\omega = 0.3$. 
The interspecies repulsion is $g_{BI}=0.2$ and the mixture is prepared in its ground state.} 
\label{fig:16}
\end{figure} 

Within this methodology, the underlying Hilbert space truncation is designated by the used orbital configuration space $C = (D;d_B;d_I)$. 
In this notation, $D$ and $d_B$, $d_I$ refer to the number of species functions [Eq. (\ref{wfn_ansatz})] and single-particle functions [Eq. (\ref{SPF})] 
of each species. 
For our numerical simulations, a primitive basis corresponding to a sine discrete variable representation involving 500 grid points is employed. 
Note also that this sine discrete variable representation intrinsically introduces hard-wall boundary conditions which are imposed herein 
at $x_\pm = \pm 50$. 
Their location do not affect the presented results since there are no appreciable densities beyond $x_{\pm}=\pm25$. 
To infer the convergence of the many-body simulations we systematically vary the numerical configuration space $C=(D;d_B;d_I)$ and ensure that all 
observables of interest become up to a certain level of accuracy insensitive. 
We remark that all many-body simulations discussed in the main text have been performed using $C=(10;3;6)$. 

To showcase the numerical convergence we exemplarily demonstrate the behavior of the center-of-mass motion of the impurities $\langle X_I (t) \rangle$ and 
of the bosonic bath $\langle X_B(t) \rangle$ following a continuous periodic driving at $\omega_D = 1.5$ for distinct orbital configurations 
$C'=(D';d'_B;d'_I)$ in Fig. \ref{fig:16}. 
Recall that at such high driving frequencies the degree of correlations inherent in the system is maximized [Fig. \ref{fig:12}]. 
Inspecting Fig. \ref{fig:16}, it can readily seen that both $\langle X_I (t)\rangle$ [Fig. \ref{fig:16} ($a$)] and $\langle X_B (t)\rangle$ [Fig. \ref{fig:16} ($b$)] 
are adequately converged since they are insensitive to the variation of the orbital configuration space $C=(D;d_B;d_I)$. 
For instance, the maximum deviation of $\langle X_I (t)\rangle$ [$\langle X_B (t)\rangle$] between the $C=(10;3;6)$ and $C'=(10;3;4)$ in the course 
of the time-evolution is at most $0.2\%$ [$0.1\%$]. 

Moreover, we present the numerical convergence of the von-Neumann entropy during the dynamics for a continuous driving characterized by 
$\omega_D=0.3$ (resonant driving) and $\omega_D=1.15$ (fast driving). 
Note here that for $\omega_D=1.15$ the von-Neumann entropy becomes maximal, see also Fig. \ref{fig:12} (a). 
To this end, we illustrate the relative difference of $S_{VN}(t)$ calculated within the $C=(10;3;6)$ and different orbital configurations $C'=(D';d_B';d_I')$ i.e.   
\begin{equation}
\Delta S_{VN}(t)_{C,C'} =\frac{\abs{S_{VN}(t)_C -S_{VN}(t)_{C'}}}{S_{VN}(t)_C}. \label{dev_Von_Neum} 
\end{equation} 
The time-evolution of $\Delta S_{VN}(t)_{C,C'}$ is shown in Fig. \ref{fig:16} at resonant driving frequencies $\omega_D=0.3$ [Fig. \ref{fig:16} (c)] and fast drivings 
with $\omega_D=1.15$ [Fig. \ref{fig:16} (d)] for a variety of orbital configurations $C'$ and fixed $C=(10;3;6)$. 
Inspecting $\Delta S_{VN}(t)_{C,C'}$ we deduce that $S_{VN}(t)$ is converged at both $\omega_D=0.3$ and $\omega_D=1.15$. 
For instance at $\omega_D=0.3$ the deviation of $\Delta S_{VN}(t)_{C,C'}$ with $C=(10;3;6)$ and $C'=(10;2;6)$ [$C=(8;2;6)$] is smaller than $1\%$ [$4\%$] throughout 
the evolution [Fig. \ref{fig:16} (c)]. 
Turning to $\omega_D=1.15$ [Fig. \ref{fig:14} (d)], we observe that $\Delta S_{VN}(t)_{C,C'}$ between the orbital configurations 
$C=(10;3;6)$ and $C'=(10;2;6)$ [$C'=(8;2;6)$] acquires a maximum value of the order of $1\%$ [$5\%$] in the course of the time-evolution. 
Additionally, let us comment that the same analysis has been done for all other observables and driving frequencies discussed in the main text 
and found to be sufficiently converged as well (results not shown here for brevity). 
\begin{figure}[ht]
\centering
\includegraphics[width=0.4\textwidth]{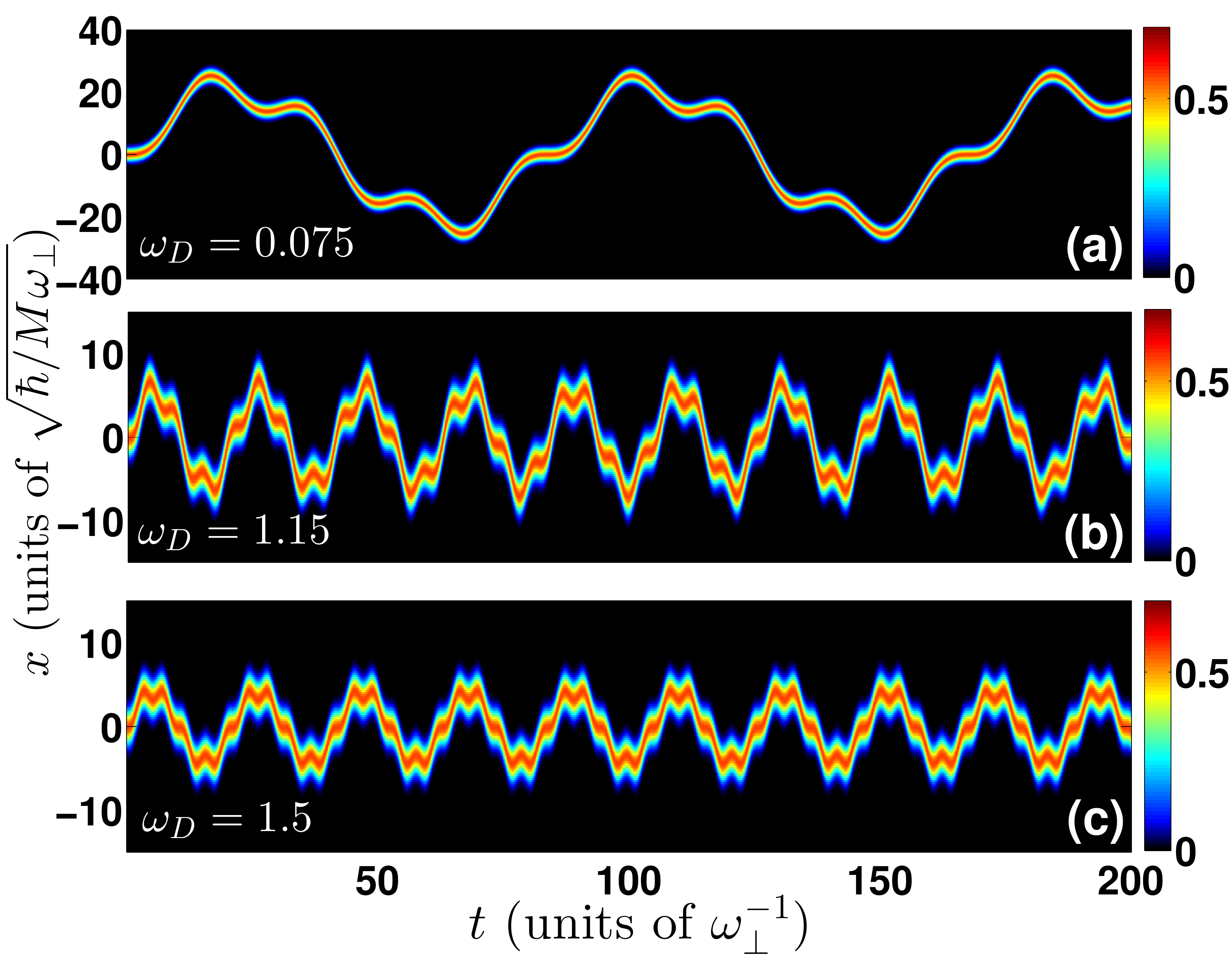}
\caption{Time-evolution of the single-particle density of two bosons trapped in a continuously shaken harmonic oscillator potential 
for specific driving frequencies $\omega_D$ (see legends). 
The system consists of $N=2$ repulsively interacting bosons with $g=0.4$. 
It is trapped in a harmonic oscillator with frequency $\omega=0.3$ and it is initialized into its ground state.} 
\label{fig:18}
\end{figure}

\section{Shaking dynamics of two-bosons}\label{shaking_single_component} 

To expose the effects caused by the presence of the bosonic bath on the dynamical response of the impurities described in the main text, we briefly 
discuss the dynamics of two bosons trapped in a continuously shaken harmonic trap. 
In particular, we consider two ($N=2$) repulsively interacting bosons in a harmonic trap of frequency $\omega = 0.3$. 
The system is initialized into its ground state with interparticle interaction strength $g= 0.4$. 
To induce the dynamics the harmonic trap is periodically shaken throughout the time-evolution and the system obeys the following Hamiltonian  
\begin{equation}\label{Hamiltonian_single_component}
 \begin{split}
&H =\sum_{i = 1}^{N} -\frac{\hbar^2}{2M}\bigg(\frac{\partial}{\partial x_i}\bigg)^2  
+\sum_{i = 1}^{N} \frac{1}{2} M \omega^2 x_i^2\\&+\frac{1}{2}M\omega^2 \sum_{i=1}^N\big(x_i - \mathcal{A}\sin(\omega_D t)\big)^2 + g\sum_{ i \geq j }^{} \delta(x_i - x_j).
\end{split}
\end{equation} 
In this expression, $\mathcal{A}$ and $\omega_D$ refer to the amplitude and the frequency of the driving respectively. 
To perform a direct comparison with the observations made in Section \ref{continuous} we use $\mathcal{A}=20$ and study the dynamics for different driving frequencies $\omega_D$ 
while keeping fixed all other parameters of the system. 

The resulting time-evolution of the two boson single-particle density, $\rho^{(1)}(x;t)$ is illustrated in Fig. \ref{fig:18} for different driving frequencies. 
As it can be seen in Fig. \ref{fig:18} (a), for small driving frequencies such as $\omega_D = 0.075$ the two bosons follow their driven potential 
and undergo an overall oscillatory motion. 
At the initial evolution times $\rho^{(1)}(x;t)$ moves to the $x>0$ direction reaching $(x\approx27)\gg (\mathcal{A}=20)$ at $t\approx16$ and then 
turns towards $x<0$ due to the presence of the trap while featuring a backward motion around $t\approx27$. 
Subsequently $\rho^{(1)}(x;t)$ moves to the $x<0$ direction performing a similar to the above-mentioned backward and forward motion until it arrives at $x\approx -27$ 
where it again turns its motion to the trap center. 
In this way, the first oscillation period of the driving is completed and afterwards a similar to the above-described motion occurs within 
each driving cycle [Figs. \ref{fig:18} (a)]. 
Note that for these weak driving frequencies the two-boson dynamical response is similar to the one of the two impurities immersed in a BEC background, 
compare Fig. \ref{fig:8} ($b_1$) and Fig. \ref{fig:18} (a). 
A notable difference occurring in the response of the aforementioned setups is that the shape of the single-particle density of the two impurities changes in the course 
of the time-evolution due to their collisions with the BEC medium, see e.g. Fig. \ref{fig:8} ($b_1$) at $t\approx34$, an event that is absent in the dynamics of the two bosons. 
We remark that a similar overall phenomenology regarding the shaken dynamics of two impurities inside a BEC and the two bosons takes place also for driving frequencies $\omega_D<0.6$. 

However, entering the driving regime with $\omega_D>0.5$ significant alterations between the responses of these setups occur. 
To exemplify these differences we showcase $\rho^{(1)}(x;t)$ of two bosons for $\omega_D=1.15$ and $\omega_D=1.5$ in Figs. \ref{fig:18} (b) and (c) respectively. 
Indeed, in both cases $\rho^{(1)}(x;t)$ performs a multifrequency oscillatory behavior of constant amplitude. 
This is in sharp contrast to the time-evolution of two impurities shown in Figs. \ref{fig:8} ($b_3$), ($b_4$) where $\rho^{(1)}_{I}(x;t)$ exhibits 
a decaying amplitude oscillatory motion within the bosonic bath throughout the dynamics. 
Also, $\rho^{(1)}_{I}(x;t)$ due to impurity-BEC interactions shows a spatially delocalized behavior for $t>100$ [Fig. \ref{fig:8} ($b_3$), ($b_4$)] whilst 
$\rho^{(1)}(x;t)$ exhibits a localized shape throughout the evolution [Fig. \ref{fig:18} ($b$), ($d$)]. 
Summarizing, we deduce that for $\omega_D>0.6$ the dynamical response of two shaken bosonic impurities is very different from the one of two bosons. 
This behavior can be explained by the fact that for a fast shaking ($\omega_D>0.6$) of the harmonic trap the impurities motion is mainly restricted 
within their host and therefore impurity-BEC interaction effects dominate the dynamics.

%%%%%%%%%%%%%%%%%%%%%%%%%%%%%%%%%%%%%%%%%%%%%%%%%%%%%%%%%%%%%%%%%%%%%%%%%%%%%
%%%%%%%%%%%%%%%%%              Bibliography                  %%%%%%%%%%%%%%%%
%%%%%%%%%%%%%%%%%%%%%%%%%%%%%%%%%%%%%%%%%%%%%%%%%%%%%%%%%%%%%%%%%%%%%%%%%%%%%
\bibliography{MB_shaking}{}
\bibliographystyle{apsrev4-1}

\end{document}